\documentclass[preprint,amsmath,amssymb,showpacs,showkeys,pre]{revtex4}
\usepackage[dvips]{epsfig}

\def\*#1{\left[{\cal #1}\right]}

\begin{document}

\title{Diffraction gratings of isotropic negative phase--velocity materials}

\author{
Ricardo Angel Depine $^{(1)}$ and Akhlesh Lakhtakia$^{(2)}$\\
$^{(1)}${\em{Grupo de Electromagnetismo Aplicado, Departamento de
F\'{\i}sica, }\\
{Facultad de Ciencias Exactas y Naturales, Universidad de
Buenos Aires, }\\
{Ciudad Universitaria, Pabell\'{o}n I},
{ 1428 Buenos Aires, Argentina}}\\
$^{(2)}${\em{Computational \& Theoretical Materials Sciences Group (CATMAS), }\\
Department of Engineering Science \& Mechanics,\\
{Pennsylvania State University, University Park, PA 16802--6812, USA}}\\
}


\begin{abstract} 
{\small Diffraction of electromagnetic plane waves by the gratings made by periodically corrugating the exposed planar boundaries
of homogeneous, isotropic, linear dielectric--magnetic half--spaces is examined. The phase velocity vector
in the diffracting material can be either co--parallel or
anti--parallel to the time--averaged Poynting vector, thereby allowing for the material to be classified as of either the positive
or the negative negative phase--velocity
(PPV or NPV) type. Three methods used for analyzing dielectric gratings~---~the Rayleigh--hypothesis
 method, a perturbative approach, and  the C formalism~---~are extended here to encompass NPV gratings by a careful
consideration of field representation inside the refracting half--space. 
 Corrugations of both symmetric as well as asymmetric shapes are studied, as also the diversity of grating response
to the linear polarization states of the incident plane wave. The replacement of PPV grating by its NPV analog affects only nonspecular
diffraction efficiencies when the corrugations are shallow, and the effect on specular diffraction efficiencies intensifies as the corrugations
deepen. Whether the type of the refracting material is NPV or PPV
is shown to affect surface wave propagation as well as resonant excitation of surface waves.}
\end{abstract} 



\maketitle

\section{Introduction}

Diffraction gratings are not only exploited by nature for the 
production of color \cite{Parker2000} but have also
been extensively used in optics for several centuries 
\cite{Harrison1949}. During the last century, tremendous
progress in manufacturing techniques made diffraction gratings  as the spectral
dispersive elements of choice. Spectacular progress was also made on 
the numerical solution of the time--harmonic
Maxwell equations for diffraction gratings, thereby greatly 
facilitating analysis and design \cite{Maystre1992}. Given
such remarkable developments, one would think that the last word on 
diffraction gratings is imminent; but the emergence
of isotropic dielectric--magnetic materials exhibiting phase
velocity vector opposed in direction to the time--averaged Poynting vector \cite{Shelby2001}-\cite{Pendry2003} has opened new prospects for diffraction gratings.

Typically, a diffraction grating is a slab of either a metal or a 
dielectric material whose exposed surface is periodically
corrugated. When a plane wave is incident on this surface, it is 
reflected partially in the specular direction fixed by Snel's law 
\cite{Kwan2002} and partially
in nonspecular directions fixed by the periodicity of the corrugated 
surface in relation to the free--space wavelength.
In addition, specular as well as nonspecular refraction into the slab 
may also occur, depending on the type
of material. Finally, reflections from the back surface also 
contribute to the overall reflection from the grating,
but those need not be considered when examining the essence of the 
phenomenon of diffraction. As the angle of
incidence is changed, specular as well as nonspecular components of 
the reflected field wax and wane, which phenomenon
is technologically exploitable.

What would happen if a diffraction grating were to be made of an 
isotropic dielectric--magnetic material? If the
phase velocity and the time--averaged Poynting vectors in this 
material are co--parallel, then the effects are not
qualitatively different from a diffraction grating made of simply an 
isotropic dielectric material \cite{Lakhtakia1989}, while
the quantitative differences are due to differences in the relative 
impedance and the wavenumber inside the diffracting
material \cite{LesterDepine1994}. This paper is devoted to the case when the phase velocity 
and the time--averaged Poynting vectors in the diffracting material
are oppositely directed. Though several names have been proposed for 
this class of materials, we think that the
most descriptive is: { negative phase--velocity} (NPV) materials. In 
contrast, the
phase velocity and the time--averaged Poynting vectors are 
co--parallel in positive phase--velocity
(PPV) materials. PPV materials are, of course, commonplace.

The plan of this paper is as follows: The boundary value problem of a diffraction grating
is presented in Section II, with careful delineation of field characteristics in
the refracting half--space. Three methods of solving the boundary value problem
 are extended in Section III to encompass NPV refracting materials. These methods are: (i) the Rayleigh--hypothesis
 method, (ii) perturbative approach, and (iii) the C formalism.  Numerical results
 for corrugations of both symmetric as well as asymmetric shapes are presented in Section IV,
 and the effects of replacing a PPV material by its NPV analog, or {\em vice versa\/},
 are extracted from those results.
An $\exp(-i\omega t)$ 
time--dependence is implicit, with $i=\sqrt{-1}$,
$\omega$ as the angular frequency, and $t$ as the time.

\section{Boundary value problem}
In a rectangular coordinate system $(x,y,z)$, we consider the 
periodically corrugated
boundary $y=g(x)=g(x+d)$ between vacuum and a homogeneous, isotropic, 
linear, passive, dielectric--magnetic
material, with $d$ being the corrugation period.
The region $y>g(x)$ is vacuous, whereas the material occupying the 
region $y<g(x)$ is
characterized
by complex--valued scalars
$\epsilon_2=\epsilon_{2R} + i  \epsilon_{2I}$ and $\mu_2=\mu_{2R} + i\mu_{2I}$
that depend on $\omega$.
If this medium is of the NPV type, then the following three 
conditions hold equivalently \cite{Lakhtakia2003,DepineLakhtakia2004}:
\begin{equation}
\left.\begin{array}{ll}
(\vert\epsilon_2\vert-\epsilon_{2R})(\vert\mu_2\vert-\mu_{2R})-\epsilon_{2I}\mu_{2I} > 
0\\[5pt]
\epsilon_{2R}\mu_{2I} +\epsilon_{2I}\mu_{2R}<0\\[5pt]
\epsilon_{2R} |\mu_2| + \mu_{2R} |\epsilon_2| <0
\end{array}\right\}
\,. \label{conditrad2}
\end{equation}
None of the three conditions hold for a PPV material.

\begin{figure}[ht] 
\begin{center} 
\begin{tabular}{c}
\includegraphics[width=4in]{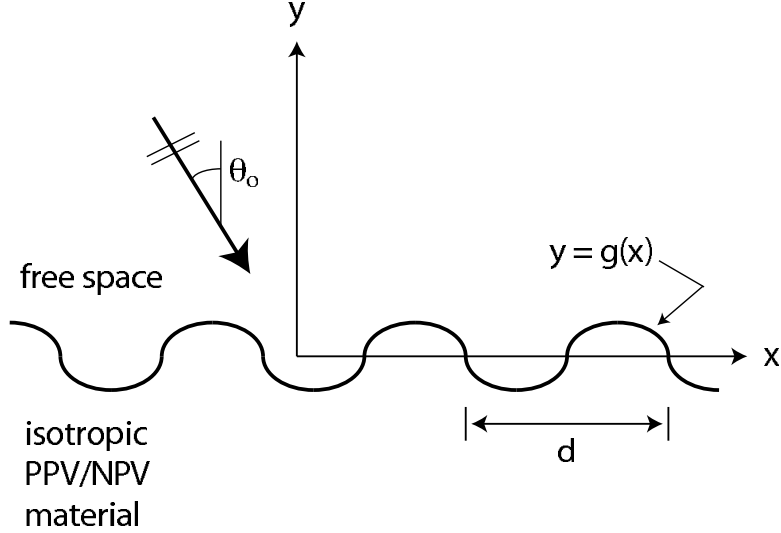} 
\end{tabular}
\end{center} 
\caption[example]{ \label{geomfig} Schematic of the boundary value problem. A plane wave is incident at an angle
 $\theta_0$, $(\vert\theta_0\vert<\pi/2)$, with 
respect to the $y$ axis on the periodically corrugated boundary $y=g(x) = g(x+d)$ between free space and a homogeneous,
isotropic, linear dielectric--magnetic material.}
\end{figure}

A linearly polarized electromagnetic plane wave is incident on this 
boundary from the region
$y>g(x)$ at an angle $\theta_0$, $(\vert\theta_0\vert<\pi/2)$, with 
respect to the $y$ axis, as shown in Figure \ref{geomfig}. This plane wave can be
either $s$--polarized or $p$--polarized \cite{BornWolf1980}. Given 
the orientation of the plane wave
with respect to the grating plane (i.e., the $xy$ plane), all 
reflected and transmitted fields must be linearly polarized
in the same way as the incident plane wave \cite{Waterman1975}.

\bigskip
\bigskip

Let the function $f(x, y)$ represent the $z$--directed component of 
the total electric
field for the $s$--polarization  case, and the $z$--directed component
of the total magnetic field for the $p$--polarization case.
After starting from the time--harmonic Maxwell equations,  this 
function can be shown to be a solution of the Helmholtz equations
\begin{equation}
\left.\begin{array}{ll}
\Big(\nabla^2 + \frac{\omega^2}{c^2}\Big) f(x,y)=0\,,&\qquad y>g(x)\\[5pt]
\Big(\nabla^2 + \frac{\omega^2}{c^2}\epsilon_2\mu_2\Big) 
f(x,y)=0\,,&\qquad y<g(x)
\end{array}\right\}\,,
\label{helmholtz1}
\end{equation}
where  $c$ is the speed of light in free space (i.e., vacuum).
Outside the corrugation region ${\rm max}\,g(x)>y>{\rm min}\,g(x)$, the field
$f(x, y)$ is rigorously represented as a superposition of plane waves as follows:
\begin{equation}
f(x,y) =
\left\{
\begin{array}{ll}
\exp\left[ i\,( \alpha_{0} x - \beta^{(1)}_0 y)\right] +  \\[5pt]
\qquad+\sum_{n=-\infty}^{\infty} c_n^{(1)} \,\exp\left[i\,( \alpha_n 
x + \beta^{(1)}_{n} y)\right] \,, &\qquad
y > \mbox{max}\,g(x)\\[5pt]
\sum_{n=-\infty}^{\infty} c_n^{(2)} \,\exp\left[i\,( \alpha_n x - 
\beta^{(2)}_{n} y)\right] \,, &\qquad
y < \mbox{min}\,g(x)
\end{array}\right.\,.
\label{expansion}
\end{equation}
Here, $c_n^{(p)}$, ($-\infty< n< \infty$; $p=1,\,2$), are scalar 
coefficients to be determined
by the solution of a boundary value problem, while
\begin{equation}
\left.
\begin{array}{ll}
\alpha_n = \frac{\omega}{c}\, \sin\theta_0 + {2n\pi}/{d}\\
\beta^{(1)}_n = \sqrt{\frac{\omega^2}{c^2}  - \alpha_n^2} \\[5pt]
\beta^{(2)}_n = \sqrt{\frac{\omega^2}{c^2}\epsilon_2\,\mu_2 - \alpha_n^2}
\end{array}\right\}\,. \label{alfabeta}
\end{equation}
Both $\beta_n^{(1)}$ and  $\beta_n^{(2)}$ are double--valued functions by virtues of their
definitions as square--roots. If $\beta_n^{(p)}$ represents an {\em upgoing\/} plane wave, then $-\beta_n^{(p)}$
represents a {\em downgoing\/} wave, and {\em vice versa\/}. We must choose the correct signs for  all $\beta_n^{(1)}$
as upgoing plane waves as well as for  all $\beta_n^{(2)}$ as downgoing plane waves.
If $\beta_n^{(p)}$ is real--valued, it corresponds to a propagating plane wave; otherwise, it indicates evanescence.

Note that $\beta^{(1)}_n$ is either purely real
or purely imaginary; and the condition
\begin{equation}
\left.\begin{array}{ll}
{\rm Re} \left[\beta^{(1)}_n\right]  \geq 0\\[5pt]
{\rm Im} \left[\beta^{(1)}_n\right]  \geq 0
\end{array}\right\}\, \quad\forall n
\end{equation}
is appropriate for upgoing plane waves in the vacuous half--space $y>g(x)$. As the direction and/or the angular
frequency of the incident plane wave change,
$\beta^{(1)}_n$ may change from purely real to purely imaginary, or {\em vice versa\/}. Such alterations
are usually marked by noticeable discontinuities in the diffraction spectrums and, hence,
are called {\em Rayleigh--Wood anomalies} although their occurences are not at all anomalous \cite{Maystre1992}.

The refracting half--space $y < g(x)$ being filled by
a material medium, $\epsilon_{2I}>0$ and $\mu_{2I} >0$ by virtue of causality \cite{BohrenHuffman1983, Hu1989}.
The refracted plane waves must attenuate as $y\to-\infty$, which 
requirement leads to the condition
\begin{equation}
  {\rm Im}\left[\beta^{(2)}_n\right] > 0\, \quad \forall n\,. \label{imbetan2}
  \end{equation}
This condition on ${\rm Im}\left[\beta^{(2)}_n\right]$ automatically fixes the sign of
${\rm Re} \left[\beta^{(2)}_n\right] $, regardless of the signs of 
$\epsilon_{2R}$ and $\mu_{2R}$; furthermore,
  the transformation $\left\{ \epsilon_{2R}\to-\epsilon_{2R},
\mu_{2R}\to-\mu_{2R}\right\}$
alters the signs of the real parts of all $\beta_n^{(2)}$.

In order to find the scalar coefficients $c_n^{(p)}$,  we must
apply the boundary conditions at  $y=g(x)$. These conditions,
expressing the continuity of the tangential components of the total 
electric field  and the total magnetic field, can be written
as
\begin{equation}
\left.\begin{array}{ll}
f(x,g(x)+) = f(x,g(x)-)\\[5pt]
\hat{n}\cdot\nabla f(x,g(x)+)=\sigma^{-1}\,\hat{n}\cdot\nabla f(x,g(x)-)
\end{array}\right\}\,,
  \label{bcs}
\end{equation}
where $\sigma=\mu_2$ for the $s$--polarization case and 
$\sigma=\epsilon_2$ for the $p$--polarization case,
while $\hat{n}$ is a unit vector normal to the boundary $y=g(x)$.

\section{Methods of Solution}
Analytical solution of the stated boundary value problem is well nigh 
impossible, except
in a perturbative sense. Early numerical techniques \cite{vandenBerg1981} relied on the so--called Rayleigh hypothesis,
according to which the expansions (\ref{expansion}) can be assumed 
valid at $y=g(x)\pm$ \cite{Rayleigh1907}. Therefore,
those techniques were not applicable for deeply corrugated boundaries 
\cite{Waterman1975}. 
The limitations of the Rayleigh hypothesis
were overcome by the T--matrix method \cite{Lakhtakia1989,ChuangKong1981} and  the C formalism \cite{Chandezonetal1980, Lietal1999}, of which the latter displays superior
performance. All of these techniques require various 
degrees of computational proficiency, and so
we applied three different methods of solution for NPV diffraction gratings~---~as checks
on each other, as applicable.

\begin{center}{\em (a) Rayleigh--hypothesis method}\end{center}
This method was enunciated by Rayleigh \cite{Rayleigh1907} for gratings made by sinusoidally
corrugating a perfectly reflecting sheet. According to his hypothesis, the
expansions (\ref{expansion}) have to be used in the boundary conditions (\ref{bcs}).
Both resulting equations are then projected into the Rayleigh basis
$\left\{\exp(i \, \alpha_mx)\right\}_{m=-\infty}^{\;\;\;\;+ \infty}$, 
in order to
obtain a linear system of equations for all $c_n^{(p)}$. The 
refraction coefficients $c_n^{(2)}$
are then eliminated \cite{Maradudin1982,LesterDepine1996} to yield

\begin{eqnarray}
&&\sum_{n=-\infty}^{\infty} \frac{(1-\sigma)\,
\left[ \beta^{(1)}_{n} \beta^{(2)}_{m}+\alpha _n\alpha_m \right]
-\frac{\omega ^2}{c^2} \left[ \mu_2 \epsilon_2 - \sigma  \right]}
{\beta^{(2)}_m - \beta^{(1)}_n}
D_{mn}\Bigl(\beta^{(1)}_n - \beta^{(2)}_m \Bigr)\, c_n^{(1)} = \nonumber \\
&&\frac{(1-\sigma)\,
\left[ \beta^{(1)}_{0} \beta^{(2)}_{m}-\alpha _0\alpha_m \right]
+\frac{\omega ^2}{c^2}\left[ \mu_2 \epsilon_2 - \sigma  \right]}
{\beta^{(2)}_m + \beta^{(1)}_0}
D_{m0}\Bigl(-\beta^{(1)}_0 - \beta^{(2)}_m \Bigr)\, \,
  \label{rosubn1}
\end{eqnarray}

for all $m$, where

\begin{equation}
\label{Dmn}
D_{mn}(u) = \frac{1}{d}\,\int_{0}^{d} \exp{[- i \, \frac{2\pi}{d}
(m-n)\, x + iu
g(x) ]} \, dx \,.
\end{equation}
The summation on the left side of (\ref{rosubn1}) has to be 
appropriately truncated, and the
equations are then put in the form of a matrix equation which can be 
solved by standard numerical methods \cite{Strang1986}.

The Rayleigh hypothesis is valid when the corrugations are not deep, 
and the limit of its applicability for
sinusoidal gratings has been rigorously established. 
Millar \cite{Millar1971} showed that
the Rayleigh hypothesis is applicable for perfectly reflecting gratings of sinusoidal shape with maximum
slope not exceeding $0.448$. This limit was validated by Hill \& Celli 
\cite{HillCelli1978}, who also
noted that the methods exploiting the Rayleigh hypothesis could yield acceptable results for steeper gratings.
Depine \& Gigli \cite{DepineGigli1994} carried out detailed numerical studies to show
that the Rayleigh hypothesis can be considered adequate for dielectric sinusoidal gratings
with maximum slopes as high as $\sim 0.92$.

\begin{center}{\em (b) Perturbative approach}\end{center}
A perturbative approach applies well when the corrugations are very shallow.
The integrals $D_{mn}(u)$ can be stated exactly as the power series
\begin{equation}
D_{mn}(u)=\sum_{j=0}^\infty
\frac{i^j}{j!}\,u^j \tilde{g}^{(j)}(m-n),  \label{Dmnserie}
\end{equation}
where
\begin{equation}
\tilde{g}^{(j)}(m)= \frac{1}{d}\,\int_{0}^{d}
\,\left[g(x)\right]^j\, \exp(-im \frac{2\pi}{d}x) \,dx
\end{equation}
is the $m$-th Fourier coefficient of the function
$[g(x)]^j$. 
These coefficients can be obtained through the recurrence relation
\begin{equation}
\tilde{g}^{(j)}(m)=\sum_n\tilde{g}^{(j-1)}(m-n)\,\tilde{g}^{(1)}(n),
\,\,\,\,j \ge 1\,,
\end{equation}
beginning with
\begin{equation}
\tilde{g}^{(0)}(m)=\delta _{m0}\,,
\end{equation}
where $\delta_{mn}$ is the Kronecker delta.

Assuming the expansion \cite{LesterDepine1996,Lopez1978}
\begin{eqnarray}
c_n^{(1)}=\sum_{j=0}^\infty \frac{(-i)^j}{j!}\,c_n^{(1,j)}\,,  \label{roserie}
\end{eqnarray}
we arrive at an iterative scheme, whereby the coefficient
$c_n^{(1,j)}$, $j \ge 1$,
can be obtained in terms of all lower--order coefficients
$c_n^{(1,j-1)}$, $\cdots$, $c_n^{(1,0)}$ as follows:
\begin{eqnarray}
c_n^{(1,j)}&=&\frac{1}{M_{nn}}\Biggl\{\,N_n \Bigl[ \beta
_0^{(1)}+\beta _n^{(2)}\Bigr]^j \,
\tilde{g}^{(j)}(n) \,- \nonumber \\
&&\sum_{m} \left[ M_{nm} \sum_{q=1}^j
\binom{j}{q} \Bigl[ \beta _n^{(2)}-\beta _m^{(1)}\Bigr]^q
\,\tilde{g}^{(q)}(n-m)\,
c_m^{(1,j-q)}\right] \Biggr\}\, .  \label{ronj}
\end{eqnarray}
This scheme commences with
\begin{eqnarray}
c_n^{(1,0)}=\frac{\sigma\beta_n^{(1)}-\beta_n^{(2)}}{\sigma
\beta_n^{(1)}+\beta _n^{(2)}}
\,\delta_{n0}\,,  \label{ron0}
\end{eqnarray}
which is the planewave reflection coefficient for a perfectly flat
boundary (i.e., $g(x)\equiv 0$), and  requires the computation of
\begin{equation}
N_n=\frac{(\beta_0^{(1)}\beta_n^{(2)} - \alpha_0 
\alpha_n)(1-\sigma)+\frac{\omega^2}{c^2}
(\epsilon_2\mu_2-\sigma )}{\beta_0^{(1)}+\beta_n^{(2)}}\, \label{enesubn}
\end{equation}
and
\begin{equation}
M_{nm}=\frac{(\beta_m^{(1)}\beta_n^{(2)}+\alpha_m
\alpha_n)(1-\sigma)-\frac{\omega^2}{c^2}
(\epsilon_2 \mu_2 - \sigma )}{\beta_n^{(2)}-\beta_m^{(1)}}\, . \label{emesubnm}
\end{equation}
Provided the series (\ref{roserie}) converges rapidly, the reflection 
coefficients $c_n^{(1)}$ can be computed
quite easily even on hand--held computers.

\begin{center}{\em (c) C formalism}\end{center}
In order to avoid the use of the Rayleigh
expansions (\ref{expansion}) in the corrugation region, the C 
formalism begins with the transformation
\begin{equation}
v=y-g(x)\,.\label{change1}
\end{equation}
Accordingly, the Helmholtz equations (\ref{helmholtz1}) change to
\begin{equation}
\Bigl [
\frac{\partial^2}{\partial x^2}
-2 \dot{g} \frac{\partial^2}{\partial x \partial v }
-\ddot{g} \frac{\partial}{\partial v }
+(1+\dot{g}^2) \frac{\partial^2}{\partial v^2 }
+\frac{\omega^2}{c^2}\epsilon(v)\,\mu(v)\,\Bigr ] f(x,v)=0\,, 
\label{helmholtz2}
\end{equation}
where
\begin{equation}
\epsilon(v) = \left\{\begin{array}{ll}1\\ 
[5pt]\epsilon_2\end{array}\right.\,,\qquad
\mu(v) = \left\{\begin{array}{ll}1\\ [5pt]\mu_2\end{array}\right.\,,\qquad
v\,\left\{\begin{array}{ll} >0\\ [5pt] <0\end{array}\right.\,
\end{equation}
and
\begin{equation}
\dot{g}=\frac{dg}{dx}\,,\qquad \ddot{g}=\frac{d^2g}{dx^2}\,.
\end{equation}

Because the coefficients of the differential equation 
(\ref{helmholtz2}) depend on $v$ in a piecewise fashion, the 
$v$--dependence of $f$ is of the form $\exp\left(i \rho v\right)$ in each of the two
pieces $v>0$ and $v<0$. Following references \cite{Lietal1999} and \cite{LiChandezon1996}, we expressed the $x$--dependences of 
$\dot{g}$, $f$ and $\partial f/\partial v$ in terms of  Fourier 
series, and obtained the following matrix equation:\\

\begin{equation}
\label{eigensystem}
\left[
\begin{array}{c|c}
-\left[{{\cal B}^{(p)}}\right]^{-2} \left(\*A\*{\dot G}+\*{\dot G}\*A\right) &
\left[{{\cal B}^{(p)}}\right]^{-2}\left(\*I+\*{\dot G}\*{\dot G}\right) \\ \\ \hline\\
\*I & \*O
\end{array}\right]\,
\left[\begin{array}{c} \*F\\ \\ \hline\\
\left[{\tilde{\cal F}}\right]\end{array}\right]
=\rho^{-1}\,
\left[\begin{array}{c} \*F\\ \\ \hline\\
\left[{\tilde{\cal F}}\right]\end{array}\right]\,.
\end{equation}

In this equation, $\*A$ and $\left[{\cal B}^{(p)}\right]$
are diagonal matrixes formed by $\alpha_n$
and $\beta_n^{(p)}$, respectively;
the $(m, n)$ element of the Toeplitz matrix $\*{\dot G}$ is the
$(m-n)$--th Fourier coefficient of $\dot{g}$; 
$\*O$ and $\*I$ are, respectively, the null
and the identity matrixes; while $\*F$ and $\*{\tilde F}$ are column vectors
formed by the Fourier coefficients of $f$ and $-i \partial f/\partial v$,
respectively. Clearly, $\rho^{-1}$ is an eigenvalue of the 2$\times$2 block 
supermatrix on the left side of (\ref{eigensystem}); and the eigenvalue spectrum
of this supermatrix has to be determined for the regions  above
($p=1$) and below ($p=2$) the corrugated  surface $v=0$.

For numerical solution, the infinite system in (\ref{eigensystem}) 
must be truncated.
If only $N$ terms are kept in each Fourier series, each block in the 
$2 \times 2$ block supermatrix is a
$N \times N$ matrix, thus resulting in $2N$ eigenvalues for each 
value of $p$. In each region, all 
eigenvalues not satisfying  the radiation condition at infinity 
should be discarded in the representation
of the diffracted field. Accordingly, for $p=1$, only those 
eigenvalues are acceptable for which either $\rho$ is real--valued and
positive
or $\rho$ is complex--valued with positive imaginary part
\cite{Lietal1999,LiChandezon1996,Chandezonetal1982}.
Similarly, for $p=2$, when the region $v<0$ is filled with a PPV material,
acceptable values of $\rho$ must be either real--valued and
negative or complex--valued with negative imaginary part. However,
when the region $v<0$ is filled with a NPV material,
acceptable values of $\rho$ must be either real--valued and
positive or complex--valued with negative imaginary part.
Actual diffracting materials must be dissipative;
hence, whether the region $v<0$ is occupied by a PPV or
a NPV material, the
criterion
\begin{equation}
{\rm Im}\left[\rho\right] < 0\, \label{imrho}
\end{equation}
suffices for $p=2$.
This criterion for the eigenvalues, together with the criterion 
(\ref{imbetan2}) for selecting $\beta^{(2)}_n$, are the central modifications that
we have incorporated in the conventional C formalism for making it applicable to 
diffraction by either PPV or NPV corrugated half--spaces.

For PPV materials, Chandezon {\em et al.} \cite{Chandezonetal1982} have shown numerically and Li \cite{Li1998} 
has shown analytically that the real--valued eigenvalues and the lower--order
complex--valued eigenvalues of (\ref{eigensystem}) converge to $\pm \beta^{(2)}_n$ as 
$N$ increases. This property must also hold for NPV materials, since, as noted by Li \cite{Lietal1999}, a plane wave is an eigensolution of  (\ref{helmholtz1}) and the 
transformation (\ref{change1}) does not change the relevant eigenvalue, whether  the refracting 
material is of the PPV or the NPV type. Indeed, (\ref{eigensystem}) indicates that 
when the signs of both $\epsilon_{2R}$ and $\mu_{2R}$ are changed, the eigenvalues in 
the refracting half--space transform into their own complex conjugates. 
This is demonstrated by the sample results presented in Table 1. 

\begin{center}
\begin{tabular}{|c|c|c|c|c|}
\hline
$c\,\rho/\omega$  &$c\,\rho/\omega$ &$c\,\beta^{(2)}_n/\omega$ &$n$   \\
    with $N=11$     &   with $N=25$     &                           &     \\
\hline
$ 2.43796\mp i0.14356$ &$2.43796\mp i0.14356$    &$2.43796\pm i0.14356$    &$0$    \\
$-2.43796\pm i0.14356$ &$-2.43796\pm i0.14356$   &$-2.43796\pm i0.14356$   &$0$    \\
$2.11442\mp i0.16492$  &$ 2.10547\mp i0.16623$   &$2.10547\pm i0.16623$    &$2$    \\
$-2.11442\pm i0.16492$ &$-2.10547\pm i0.16623$   &$-2.10547\pm i0.16623$   &$2$    \\
$ 1.71339\mp i0.20374$ &$ 1.71413\mp i0.20419$   &$1.71413\pm i0.20419$    &$3$    \\
$-1.71339\pm i0.20374$ &$-1.71413\pm i0.20419$   &$-1.71413\pm i0.20419$   &$3$    \\
$ 0.94258\mp i0.30709$ &$ 1.00455\mp i0.34841$   &$1.00455\pm i0.34841$    &$4$    \\
$-0.94258\pm i0.30709$ &$-1.00455\pm i0.34841$   &$-1.00455\pm i0.34841$   &$4$    \\
$-1.00099\pm i0.34382$ &$-1.03946\pm i0.33671$   &$-1.03946\pm i0.33671$   &$-5$   \\
$ 1.00099\mp i0.34382$ &$ 1.03946\mp i0.33671$   &$1.03946\pm i0.33671$    &$-5$   \\
$ 1.71661\mp i0.19773$ &$ 1.73180\mp i0.20210$   &$1.73180\pm i0.20210$    &$-4$   \\
$-1.71661\pm i0.19773$ &$-1.73180\pm i0.20210$   &$-1.73180\pm i0.20210$   &$-4$   \\
\hline
\end{tabular}

\vspace{0.3cm}

Table 1: Some eigenvalues of (\ref{eigensystem}) for a dielectric--magnetic material
($\epsilon_2=\mp 6+i0.1,\,\mu_2=\mp 1+i0.1$) computed for truncation parameters 
$N=11$ and $N=25$, in comparison with $\beta_n^{(2)}$, for $h/d =0.1$, $\omega d/c=2\pi/0.5$
and $\theta_0=15^\circ$. The first three columns span upgoing and downgoing waves, but only downgoing
waves are acceptable in the present situation. Therefore, acceptable values of $\beta_n^{(2)}$ and $\rho$ must conform
to the restrictions (\ref{imbetan2}) and (\ref{imrho}), respectively. \\ \vspace{0.3cm}
\end{center}

\vspace{0.8cm}\begin{center}----------------------------------
\end{center}

Once the foregoing changes have been incorporated, implementation of the C formalism proceeds as usual. 
As this is well--documented in the 
literature \cite{Chandezonetal1980,Lietal1999,Chandezonetal1982},  
we only  sketch the procedure here for completeness. The following 
two steps are undertaken:
\begin{itemize}
\item[(i)] First, the function
$f(x,y)$ is written as
\begin{equation}
f(x,y) =\left\{
\begin{array}{c} f^{(1)}(x,y)\\[5pt] f^{(2)}(x,y)\end{array}\right.\,,
\qquad
y\,\left\{\begin{array}{c} >g(x)\\[5pt] <g(x)\end{array}\right.\,.
\end{equation}
Here, the sectional field functions
\begin{eqnarray}
f^{(1)}(x,y)&=& \exp\left[ i\,( \alpha_{0} x - \beta^{(1)}_0 y)\right] +
\sum_{n \in {\sf U}^{(1)}} c_n^{(1)} \,\exp\left[i\,( \alpha_n x + 
\beta^{(1)}_{n} y)\right] \,+ \nonumber \\
&& \sum_{m} \exp( i\,\alpha_m x)\,\sum_{q \in {\sf V}^{(1)}} 
C_q^{(1)} f^{(1)}_{mq}\exp\left\{i \rho_q^{(1)} [y-g(x)]\right\}
\,\label{f1a}
\end{eqnarray}
and
\begin{eqnarray}
f^{(2)}(x,y) &=&
\sum_{n \in {\sf U}^{(2)}} c_n^{(2)} \,\exp\left[i\,( \alpha_n x - 
\beta^{(2)}_{n} y)\right] \,+\nonumber \\
&& \sum_{m} \exp( i\,\alpha_m x)\,\sum_{q \in {\sf V}^{(2)}} 
C_q^{(2)} f^{(2)}_{mq}\exp\left\{i \rho_q^{(2)} [y-g(x)]\right\}
\,, \label{f2a}
\end{eqnarray}
contain $C_q^{(p)}$ as unknown scalars
with the index $q$ indicating the $q$--th eigenvalue $1/\rho_q^{(p)}$
of the 2$\times$2 block supermatrix in (\ref{eigensystem}), and 
$f_{mq}^{(p)}$ are the successive
elements of the top half of the corresponding eigenvector. The set 
${\sf U}^{(p)}$ contains the indexes corresponding to physically 
acceptable
propagating plane waves, and we note that the set ${\sf U}^{(2)}$ is 
always an empty set when the refracting material is dissipative.  In 
contrast, the set ${\sf V}^{(p)}$ contains indexes corresponding to 
physically acceptable evanescent plane waves.

\item[(ii)]
Second, (\ref{f1a}) and (\ref{f2a})  are rewritten in terms of the 
variables $x$ and $v$ only and then introduced
in the boundary conditions (\ref{bcs}). A complete set of linear 
algebraic equations is thereby obtained
for the sets of the unknown scalars $c_n^{(p)}$
and $C_q^{(p)}$. The $2N$ scalars are then calculated using standard
methods \cite{Strang1986}. 
\end{itemize}

The C formalism, not invoking the Rayleigh hypothesis and therefore not limited to gratings 
with shallow corrugations, is a very efficient and versatile theoretical tool for modeling the electromagnetic responses of
diffraction gratings of arbitrary permittivity and corrugation shape. As stated in Ref. \cite{LiChandezon1996}, the most distinctive 
feature of this formalism is its virtually uniform convergence, regardless of the incident polarization state and the
permittivity of the refracting material. Originally set up for uncoated, perfectly conducting gratings in classical mounts
\cite{Chandezonetal1980}, the essence of the formalism~---~mainly, the simplicity of the coordinate 
system (\ref{change1})~---~has allowed its extension to many other 
situations. Examples include  multilayer--coated dielectric and metallic gratings \cite{Chandezonetal1982}, conical mountings 
\cite{PopovMashev1986}, nonlinear materials \cite{PopovNeviere1994}, anisotropic materials  
\cite{Harrisetal1995,Inchaussandague1996,Inchaussandague1997} 
nonhomogeneous materials \cite{Granetetal1997}, and crossed gratings 
\cite{Granet1995}. This versatility of the 
C formalism is not matched by any other rigorous method for gratings 
\cite{Maystre1992}. We found that the characteristic
features of this formalism are valid even for diffraction gratings of isotropic NPV materials.

\begin{center}{\em (d) Conservation of energy} \end{center}
Diffraction efficiencies 
\begin{eqnarray}
e_{n}^r = \frac{{\rm Re}\left[\beta^{(1)}_n\right]}{\beta^{(1)}_0}\,\vert c_n^{(1)}\vert^2 \,, \label{ern}
\end{eqnarray}
are defined
for the propagating planewave components of the reflected field in the region $y> {\rm max} g(x)$.. 
The normalized power $P_a$ transferred across one period of the corrugated boundary into the refracting
half--space $y < g(x)$ can be 
calculated by virtue of the Poynting theorem, if the fields at the surface $y=g(x)-$ are known. The principle of conservation of energy requires that
\begin{eqnarray}
P_{a}+\sum_{n \in {\sf U}^{(1)}} e_{n}^r  =1  \,, \label{prpa1}
\end{eqnarray}
with $P_a$ being completely absorbed by the refracting material.
When we implemented any of the methods of solution presented in Sections III(a)--(c), 
we checked that the condition (\ref{prpa1}) was satisfied  to an error of 10~ppm. 

\section{Results and conclusions}
Although corrugations of different shapes are used, we confined ourselves chiefly to the most   most widely
used shape:
\begin{equation}
\label{sinugrat}
g(x) = h\,\cos\left(\frac{2\pi x}{d}\right)\,.
\end{equation}
Calculations of the diffraction efficiencies
were made for many values of the geometric ratio $h/d$ and normalized periodicity $\omega d/c$,
using one or all three of the methods of solution described in Section III, as applicable. These results are
presented and discussed in Sections IV(a)--(c).
Asymmetric counterparts 
\begin{equation}
g(x)=h_1 \cos (\frac{2\pi x}{d}) + h_2 \cos (\frac{4\pi x}{d} -\gamma)\,  \label{sum1}
\end{equation}
of the symmetric gratings (\ref{sinugrat}) are addressed in Section IV(d).

\begin{center}{\em (a) Shallow gratings}\end{center}
Let us begin with gratings described by (\ref{sinugrat}).
When the boundary 
$y=g(x)$ is perfectly flat, the only non--zero reflection
coefficient is $c_0^{(1)}$.  The transformation 
\begin{equation}
\label{PPVtoNPV}
\left\{\epsilon_{2}\to-\epsilon^\ast_{2},\,\mu_{2}\to-\mu^\ast_{2}\right\}\,,
\end{equation}
which
amounts to the replacement of a NPV/PPV half--space by an analogous PPV/NPV 
half--space, changes the phase of $c_0^{(1)}$ but not 
its magnitude \cite{Lakhtakia2003}; hence, the transformation does not affect $e_0^r$ at all.

For a shallow grating, we therefore expect that the magnitude of the specular reflection
coefficient would not be greatly affected by the transformation (\ref{PPVtoNPV}), 
but the effect of the transformation should be 
unambiguously evidenced by the nonspecular diffracted orders. 
This is indeed true, as was borne out by results computed using the perturbative approach 
of Section III(b).
  
\begin{figure}[ht] 
\begin{center} 
\begin{tabular}{c}
\includegraphics[width=2.5in]{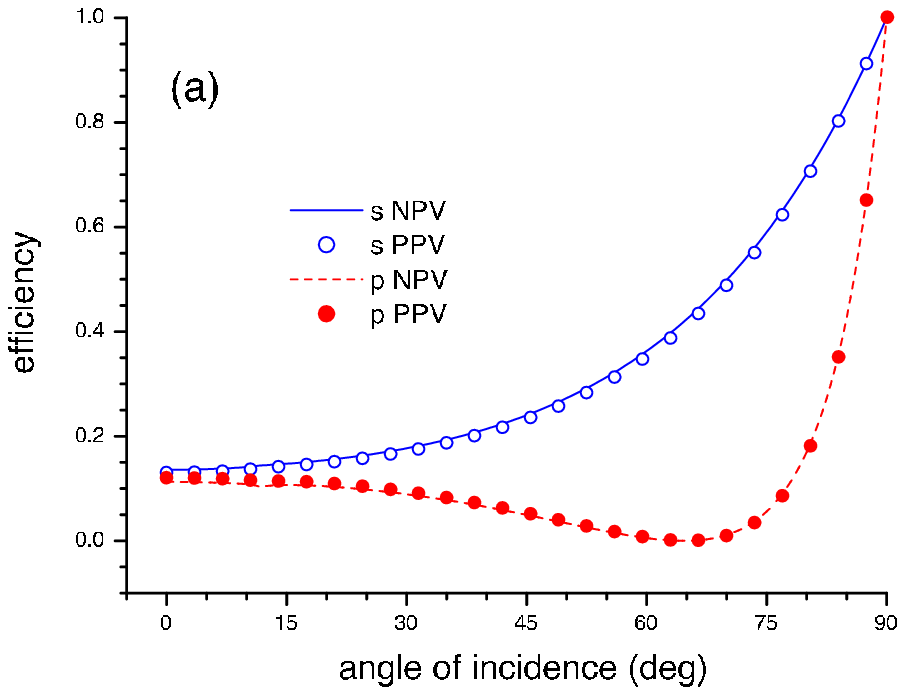} \hspace{0.3cm}
\includegraphics[width=2.5in]{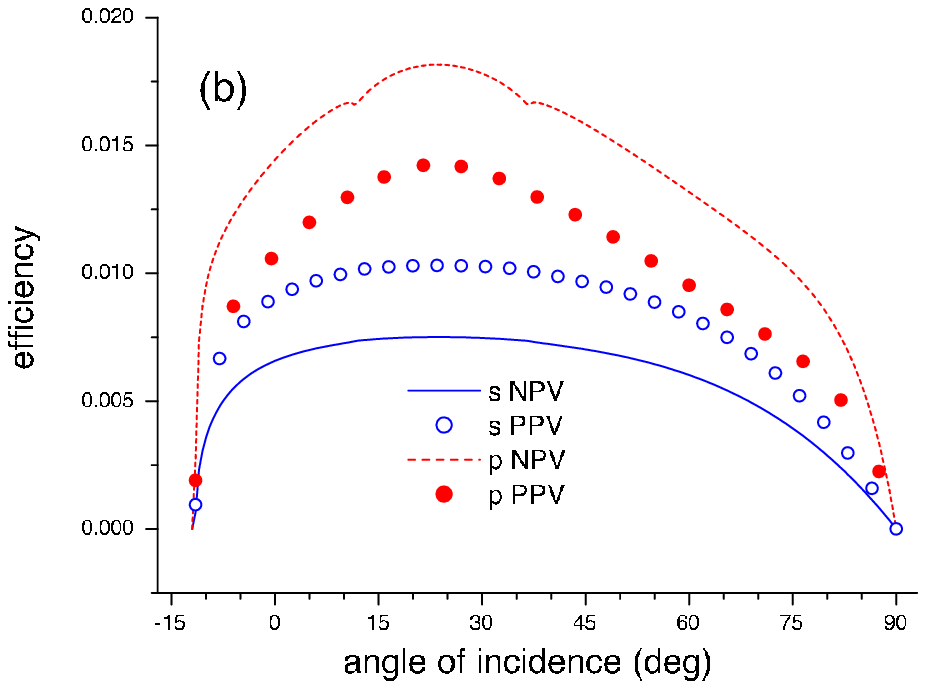} 
\end{tabular}
\end{center} 
\caption[example]{ \label{perturbresults} Diffraction efficiencies (a) $e_0^r$ and (b) $e_{-1}^r$
  as  functions of the incidence angle $\theta_0$, when
$h/d=0.07$ and $\omega d/c = 2\pi/0.8$. The refracting material is of either the
PPV ($\epsilon_2=5+i0.01,\,\mu_2=1+i0.01$) or the NPV ($\epsilon_2=-5+i0.01,\,\mu_2=-1+i0.01$)
type. Calculations were made for both the $s$-- and the $p$--polarization cases.  Note that $e_0^r(\theta_0)=e_0^r(-\theta_0)$
and that $e_{-1}^r(\theta_0)\ne 0$ only for a limited $\theta_0$--range. The same results were obtained with all three methods of solution described in Sections III(a)--(c).}
 \end{figure}

Figure \ref{perturbresults}  presents the diffraction efficiencies $e_0^r$ and $e_{-1}^r$ as 
functions of $\theta_0\in(-\pi/2,\pi/2)$ when  $h/d=0.07$ and $\omega d/c = 2\pi/0.8$.
The refracting material is of either the PPV ($\epsilon_2=5+i0.01,\,\mu_2=1+i0.01$) or the NPV 
($\epsilon_2=-5+i0.01,\,\mu_2=-1+i0.01$) type. Calculations were made for both the $s$-- and the 
$p$--polarization cases. Two Rayleigh--Wood anomalies occur at $\theta_0 \approx 11.54^\circ$ 
($\beta_1^{(1)} = 0$) and at $\theta_0 \approx 36.87^\circ$ ($\beta_{-2}^{(1)} = 0$). 

Clearly, Figure \ref{perturbresults} shows that the transformation (\ref{PPVtoNPV})
does not greatly affect $e_0^r$, except at low $\vert\theta_0\vert$. 
In contrast, the same figure shows
that the nonspecular diffraction efficiency $e_{-1}^r$,
which is non--zero only for $\sin\theta_0> -0.2$, is gravely affected by the type of the refracting material.

The diversity can be understood as follows:
When the boundary is perfectly flat, the transformation (\ref{PPVtoNPV})  leaves  
the magnitude of the reflection coefficient unchanged {\em only} for non--evanescent
incident plane waves; but that is not a true statement for incident evanescent plane waves \cite{Lakhtakia2004}.
In the troughs of a shallow grating, i.e., for ${\rm max}\,g(x)>y>g(x)$, the total field actually has both specular ($n=0$)
and nonspecular ($n\ne 0$) components, by virtue of the Rayleigh
hypothesis.  The nonspecular components are like evanescent 
plane waves because
they are characterized by ${\rm Re}\left[\beta_n^{(1)}\right]=0$. Their presence ensures that the
nonspecular reflection efficiencies, although weak for very shallow gratings,
are considerably affected~---~in contrast to the specular reflection
efficiency~---~by the transformation of the refracting material from the NPV/PPV to the PPV/NPV type.

\bigskip
 \bigskip

\begin{figure}[ht] 
\begin{center} 
\begin{tabular}{c}
\includegraphics[width=2.5in]{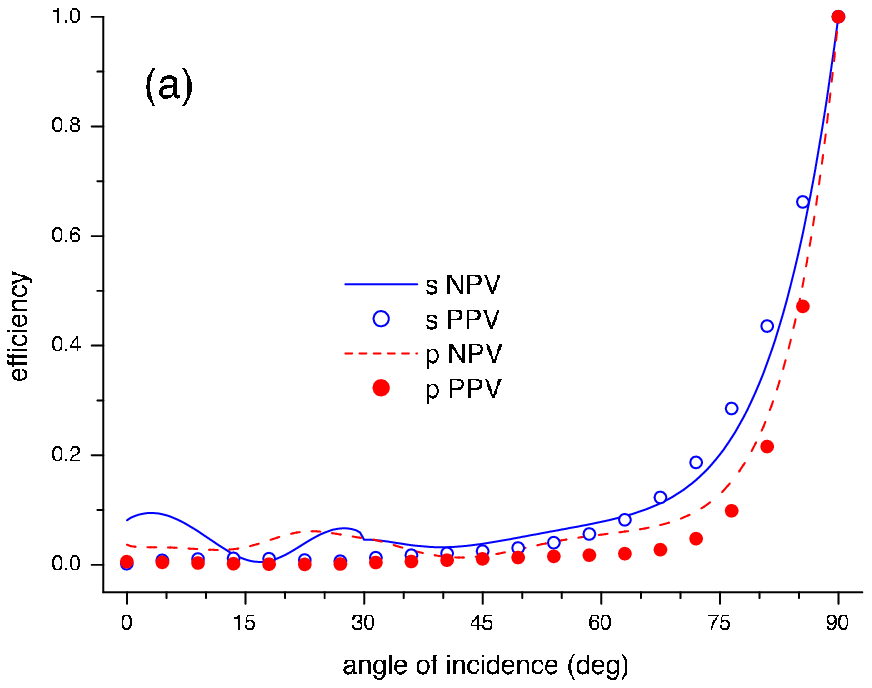} \hspace{0.3cm}
\includegraphics[width=2.5in]{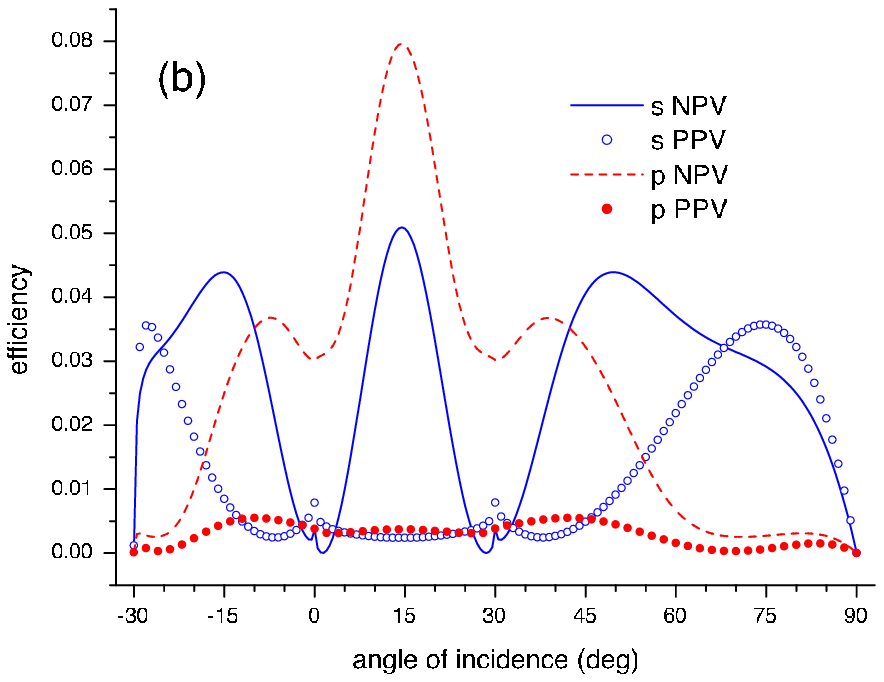} \\  \vspace{0.4cm}
\includegraphics[width=2.5in]{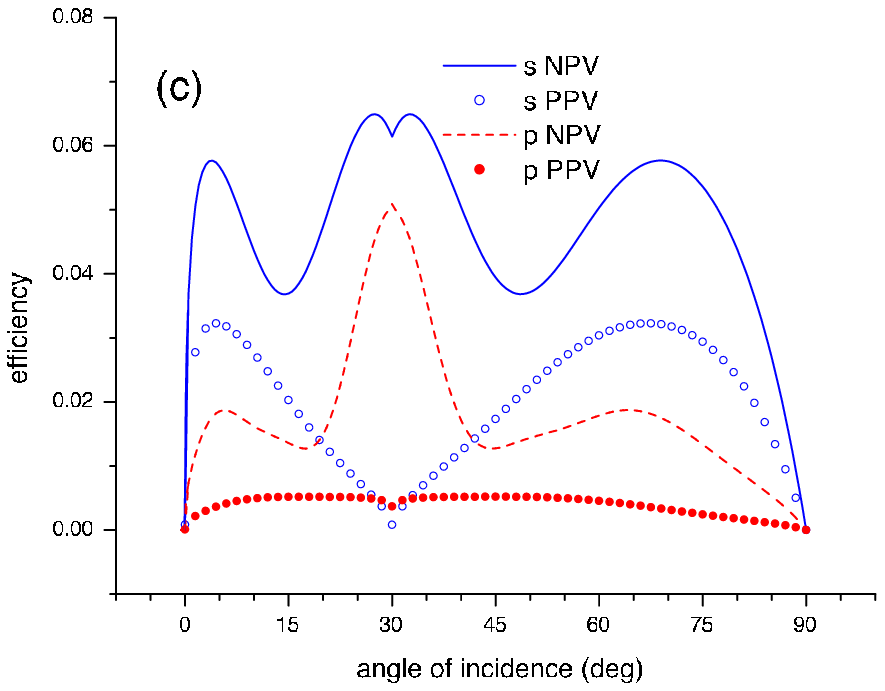} \hspace{0.3cm}
\includegraphics[width=2.5in]{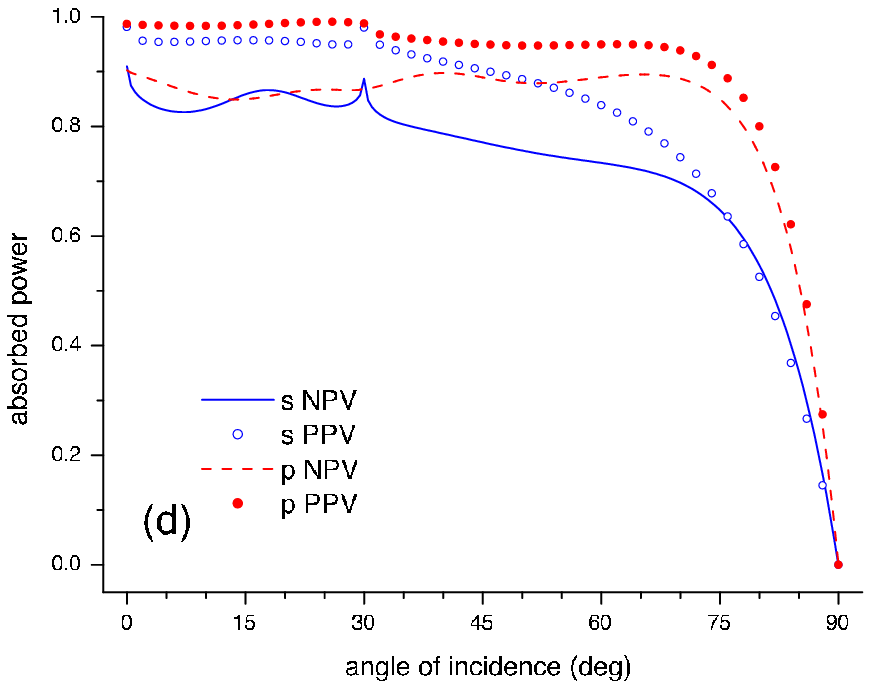} 
\end{tabular}
\end{center} 
\caption[example]{ \label{deepresults} Diffraction efficiencies (a) $e_0^r$, (b) $e_{-1}^r$, and (c) $e_{-2}^r$, and (d) normalized
absorbed power $P_a$,
  as  functions of the incidence angle $\theta_0$, when
$h/d=1$ and $\omega d/c = 2\pi/0.5$. The refracting material is of either the
PPV ($\epsilon_2=6+i0.01,\,\mu_2=1+i0.01$) or the NPV ($\epsilon_2=-6+i0.01,\,\mu_2=-1+i0.01$)
type. Calculations were made for both the $s$-- and the $p$--polarization cases, using the C formalism with
$N=29$.  Note that $e_0^r(\theta_0)=e_0^r(-\theta_0)$,
and that $e_{-1}^r(\theta_0)\ne 0$ as well as $e_{-2}^r(\theta_0)\ne 0$ only for limited $\theta_0$--ranges.}
 \end{figure}

\begin{center}{\em (b) Deep gratings} \end{center}
As the corrugations grow deeper (i.e., as $h/d$ increases in value), 
the transformation of the refracting
medium from NPV/PPV to PPV/NPV increasingly affects the specular efficiency $e_0^r$
as well. This is demonstrated by the plots of the diffraction efficiencies $e_{0,-1,-2}^r$ versus
$\theta_0$ in Figure \ref{deepresults} for $h/d=1$. As the Rayleigh hypothesis is inadequate
for sinusoidal gratings with $h/d \stackrel{>}{\approx} 0.3$ 
\cite{Inchaussandague1996}, 
the presented plots were obtained using the C formalism. 
Incidentally, Rayleigh--Wood anomalies are evident in this figure at 
$\theta_0 = 0^\circ$ ($\beta_{\pm 2}^{(1)} = 0$) and at $\theta_0 =30^\circ$ 
($\beta_{-3}^{(1)} = \beta_{1}^{(1)} = 0$).

\bigskip
 \bigskip

\newpage
\begin{center} {\em (c) Resonant surface--wave excitation}\end{center}
A mechanism that can introduce dramatic changes in the 
diffraction efficiencies when the type of the refracting material is
transformed from NPV/PPV to PPV/NPV is the resonant excitation of 
surface waves.
Surface waves are not allowed to propagate 
on a plane boundary between  vacuum and a material
whose permittivity and permeability have positive real parts (Boardman 1982).
For $p$--polarized (resp. $s$--polarized) surface waves to propagate along that 
boundary, the real part of the permittivity (resp. permeability)
of the refracting material must be negative.
Dielectric materials with negative real permittivity
are exemplified by plasmas as well as metals (Boardman 1982).
With the emergence of NPV materials, the propagation of both types
of surface waves on the same plane boundary has become possible 
\cite{Ruppin2000, Shadrivovetal2004}, although in principle for different frequencies. 

If the electromagnetic fields of the surface wave on each side of a plane boundary are described 
by (\ref{expansion}) without the term corresponding to the incident plane wave and with only the $n=0$ terms in the 
two series, dispersion relations can be easily obtained 
\cite{Ruppin2000, Shadrivovetal2004}. Thus, the wavenumber $\alpha_0$ of the surface wave satisfies the relation
\begin{equation}
\alpha_0^2=\frac{\omega^2}{c^2}\frac{\mu_2-\epsilon_2}{\mu_2^2-1}\mu_2 
\label{surfs}\,,
\end{equation}
for $s$ polarization, and
\begin{equation}
\alpha_0^2=\frac{\omega^2}{c^2}\frac{\epsilon_2-\mu_2}{\epsilon_2^2-1}\epsilon_2 
\label{surfp}\,,
\end{equation}
for $ p$ polarization. These relations apply rigorously to plane boundaries only.

To illustrate how the surface wave mechanism can affect the 
diffraction efficiencies of a grating
when the type of the refracting material is transformed from NPV/PPV to PPV/NPV, even for 
shallow corrugations,
we performed calculations with 
$\epsilon_2=-1.8+i0.01$ and
$\mu_2=1.5+i0.01$. According to the conditions (\ref{conditrad2}), this 
material is of the NPV type. 

We see, from (\ref{surfp}), that a plane boundary can
support a $p$--polarized surface wave with 
$c\,\alpha_0/\omega \approx 1.63+i 0.01$. If the transformation (\ref{PPVtoNPV}) is implemented,
the plane boundary can not support the propagation of a $p$--polarized 
surface wave; instead, as follows from (\ref{surfs}), an $s$--polarized surface wave can then
propagate with $c\,\alpha_0/\omega \approx 1.99+i 0.02$. As the real parts of both 
values of $c\,\alpha_0/\omega$ are greater than unity,
neither of the two types of surfaces waves can be resonantly excited by illuminating the plane boundary by a
 plane wave from the vacuum side. 

\begin{figure}[tb] 
\begin{center} 
\begin{tabular}{c}
\includegraphics[width=2.5in]{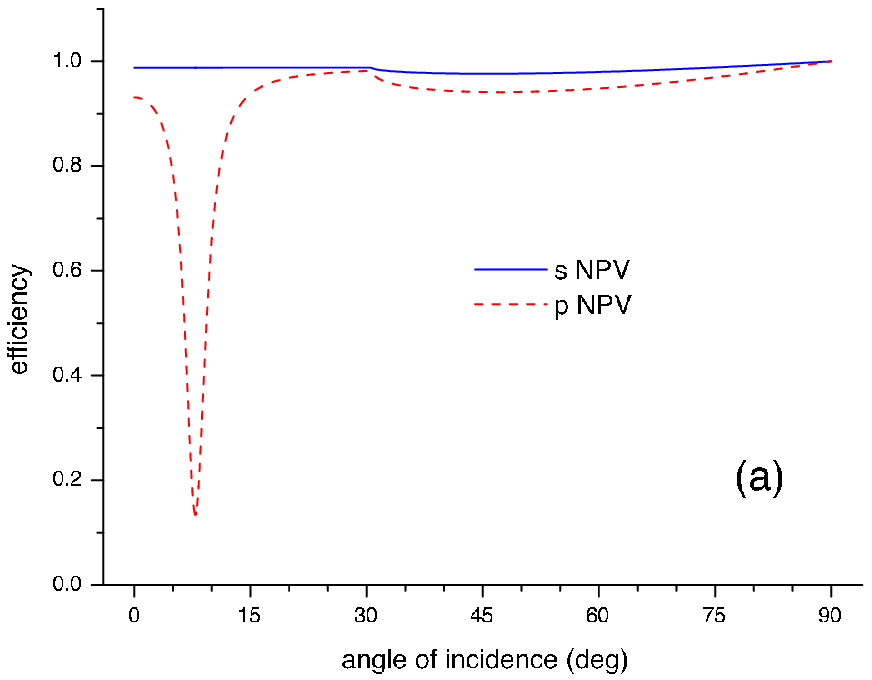} \hspace{0.3cm}
\includegraphics[width=2.5in]{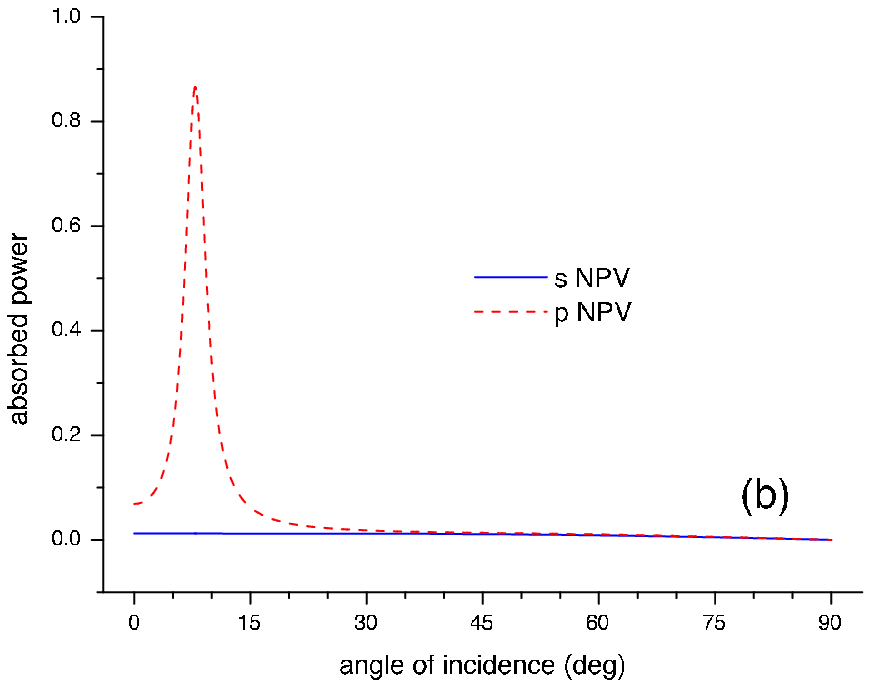} 
\end{tabular}
\end{center} 
\caption[example]{ \label{surfwave1} (a) Diffraction efficiency $e_0^r$  and (b)
normalized absorbed power $P_a$  functions of the incidence angle $\theta_0$, when
$h/d=0.07$ and $\omega d/c = 2\pi/1.51$. The refracting material is of the
NPV ($\epsilon_2=-1.8+i0.01,\,\mu_2=1.5+i0.01$) type. 
Calculations were made for both the $s$-- and the $p$--polarization cases using all three
methods presented in Section III.}
\end{figure}

However, as  is well--established in the grating literature 
\cite{Boardman1982, Raether1988},
surface waves can be coupled to propagating waves through the periodicity of a corrugated boundary.
If the period of the grating is convenently chosen, the surface wave can be coupled with 
one of the nonspecular  components (i.e., $n\ne 0$).
For example, after choosing $\omega d/c = 2\pi/1.51$ and assuming that the 
wavenumber  of the surface
wave is not appreciably altered by the corrugation, 
(\ref{alfabeta}) for $\alpha_n$ predicts a
coupling at $\theta_0 \approx 7^\circ$, when the refracting material is of the 
NPV type ($\epsilon_2=-1.8+i0.01$,
$\mu_2=1.5+i0.01$). This is confirmed by the numerical results shown 
in Figure \ref{surfwave1} for $h/d=0.07$. The zeroth--order 
efficiency curve as a function
of angle of incidence (Figure \ref{surfwave1}a) for $s$--polarization is almost flat, 
whereas for $p$--polarization it
exhibits a pronounced dip, near an angle of incidence very close to 
that predicted by the quasiplane approximation (\ref{surfp}) for surface--wave excitation. 
This dip, at $\theta_0 \approx 7.9^\circ$, is not related to a redistribution of 
the incident power between other reflected orders (i.e., a Rayleigh--Wood anomaly).
Instead, this dip is associated with a peak in the power absorbed in
the refracting material, as can be seen from the $P_a$--$\theta_0$ curve
in Figure \ref{surfwave1}b. At the dip, nearly $87\%$ of the 
$p$--polarized incident power is absorbed by the refracting material, whereas just less than 
$2\%$ of incident power is absorbed at all angles of incidence
for the other polarization case.

\bigskip
\bigskip

\begin{figure}[tb] 
\begin{center} 
\begin{tabular}{c}
\includegraphics[width=2.5in]{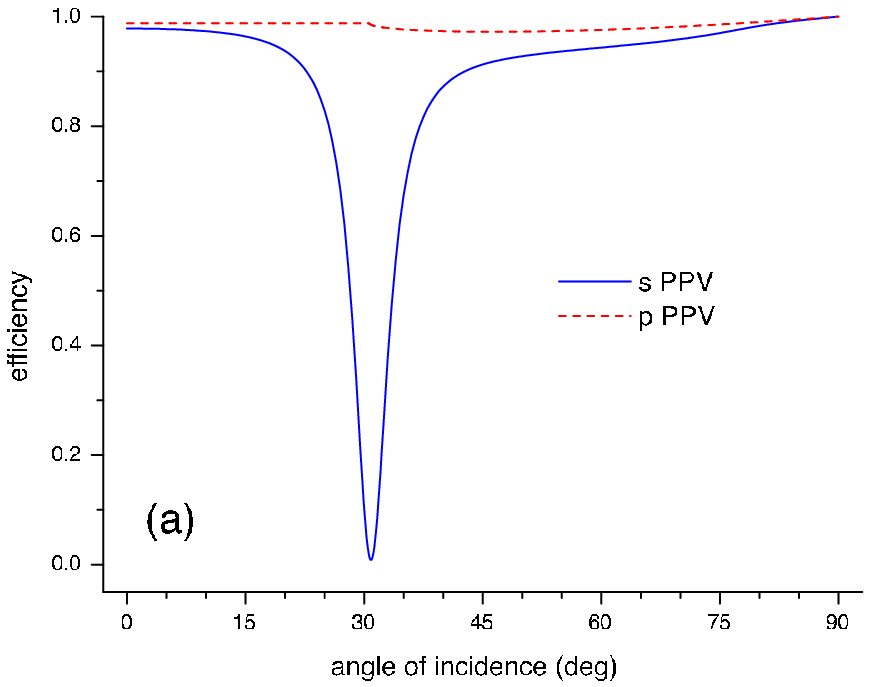} \hspace{0.3cm}
\includegraphics[width=2.5in]{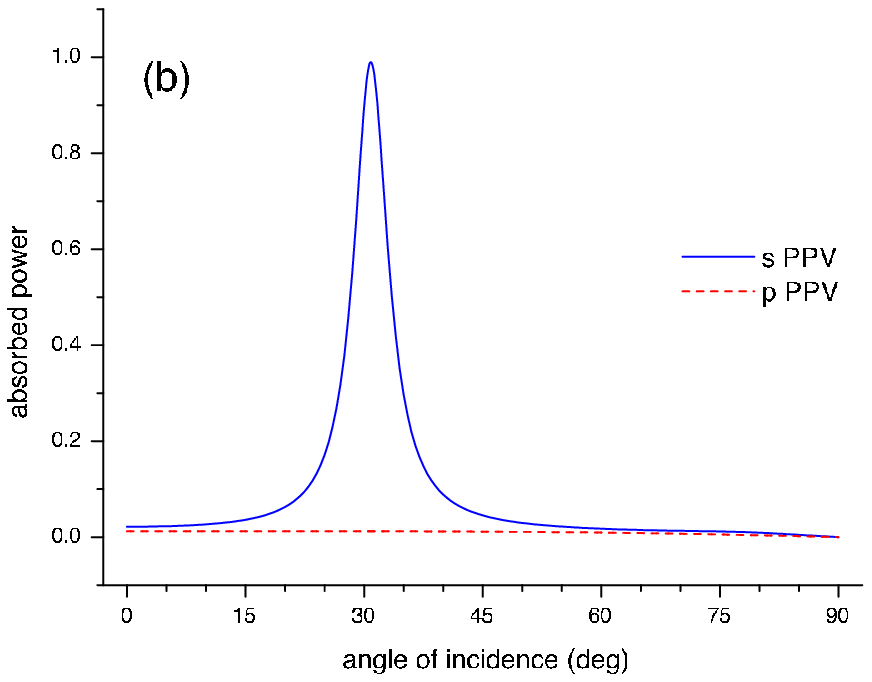} 
\end{tabular}
\end{center} 
\caption[example]{ \label{surfwave2} Same as Figure \ref{surfwave1}, except that the refracting material
is replaced by its PPV analog  ($\epsilon_2=1.8+i0.01,\,\mu_2=-1.5+i0.01$).}
 \end{figure}

That the transformation of the type of
the refracting material from NPV to PPV radically alters the conditions for surface--wave excitation is evident on comparing 
Figures \ref{surfwave1} ($\epsilon_2=1.8+i0.01$, $\mu_2=-1.5+i0.01$) and 
\ref{surfwave2} ($\epsilon_2=-1.8+i0.01$, $\mu_2=1.5+i0.01$). 
Three differences are noticeable.
First, the polarization--dependences are different: whereas the
the NPV grating  exhibits a strong absorption peak
for $p$-- but not for $s$--polarization,  its PPV analog exhibits a
strong absorption peak  for $s$-- but not for $p$--polarization.
Second, the absorption peaks occur at different angles of incidence for the two types of materials. 
The peak absorption in Figure \ref{surfwave1} occurs for $p$--polarization at $\theta_0\approx 7.9^\circ$, 
but for $s$--polarization at $\theta_0 \approx 30.9^\circ$ in Figure \ref{surfwave2}b. Third, although both 
peak absorptions are very strong, that in Figure \ref{surfwave2}, near $\theta_0 \approx 30.9^\circ$, is 
almost total (nearly $99\%$ of the $s$--polarized incident power). 

\bigskip
 \bigskip

\begin{figure}[ht] 
\begin{center} 
\begin{tabular}{c}
\includegraphics[width=2.5in]{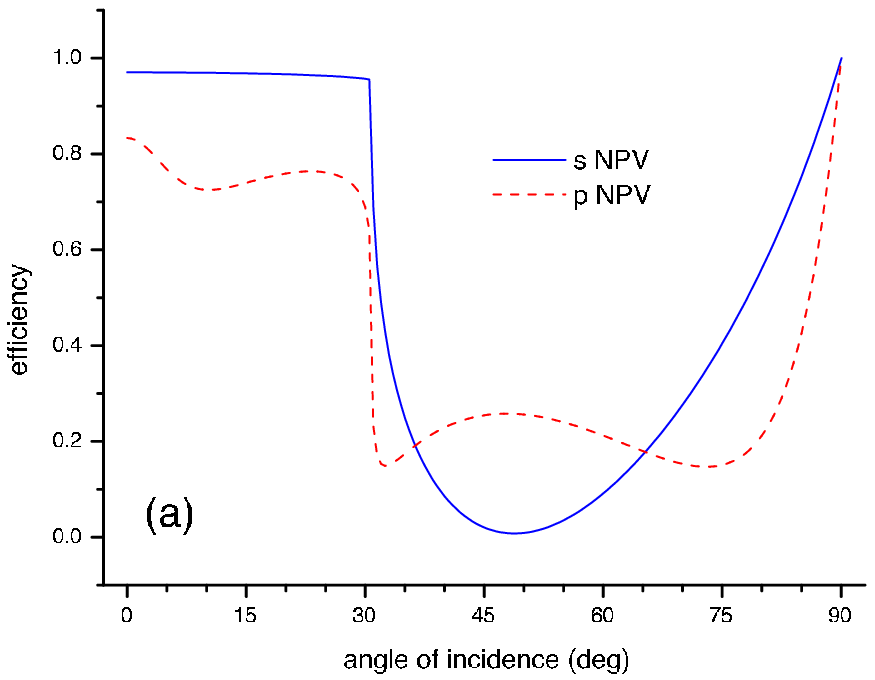} \hspace{0.3cm}
\includegraphics[width=2.5in]{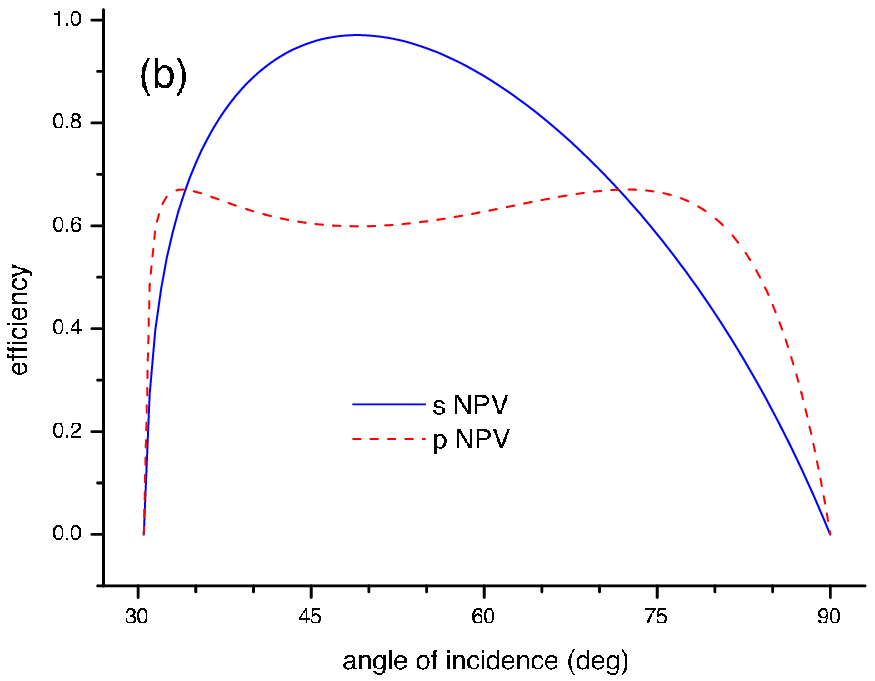} \\  \vspace{0.4cm}
\includegraphics[width=2.5in]{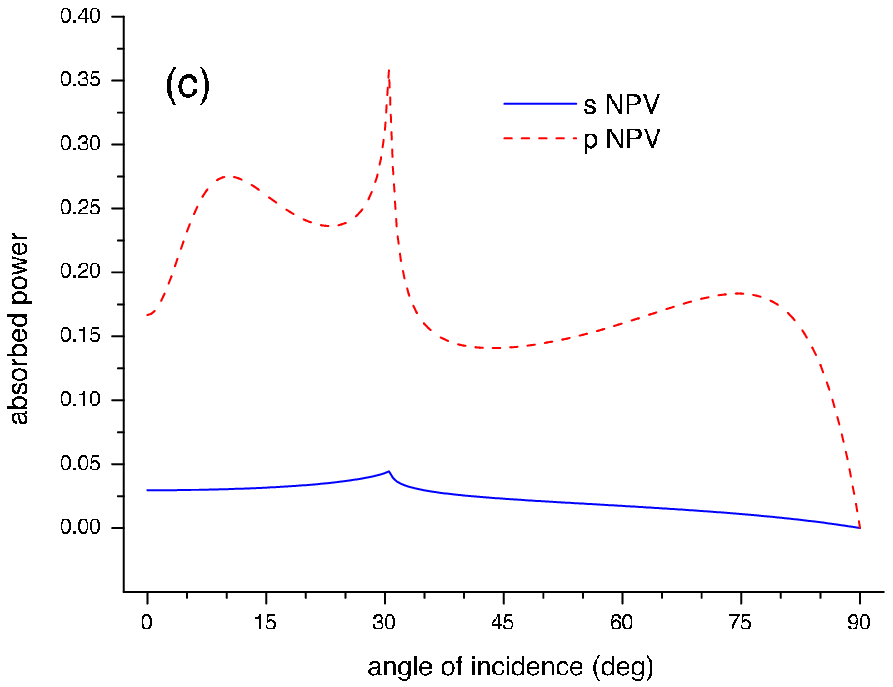}
\end{tabular}
\end{center} 
\caption[example]{ \label{surfwave3} Diffraction efficiencies (a) $e_0^r$ and (b) $e_{-1}^r$ and (c)
normalized absorbed power $P_a$ as 
functions of the incidence angle $\theta_0$, when $h/d=1$ and 
$\omega d/c = 2\pi/1.51$. The refracting material is of the
NPV ($\epsilon_2=-1.8+i0.01,\,\mu_2=1.5+i0.01$) type. 
Calculations were made for both the $s$-- and the $p$--polarization cases using the C 
method with $N=29$.}
\end{figure}

\begin{figure}[ht] 
\begin{center} 
\begin{tabular}{c}
\includegraphics[width=2.5in]{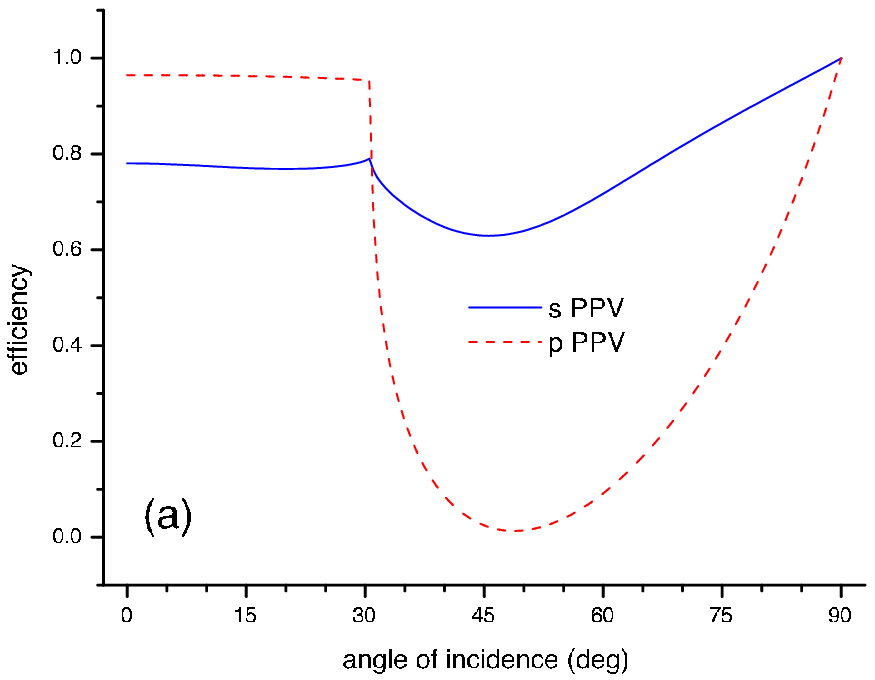} \hspace{0.3cm}
\includegraphics[width=2.5in]{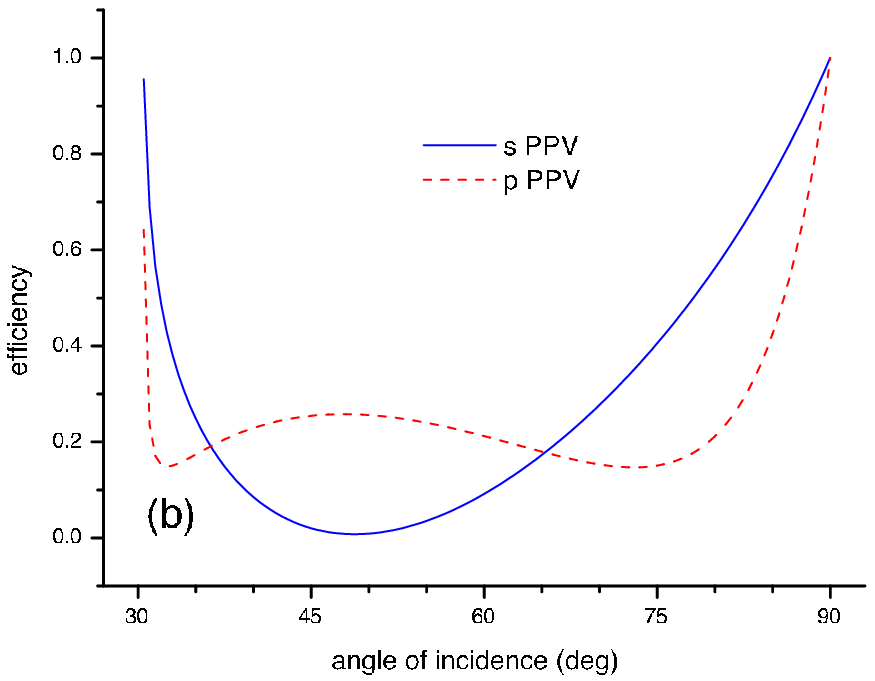} \\  \vspace{0.4cm}
\includegraphics[width=2.5in]{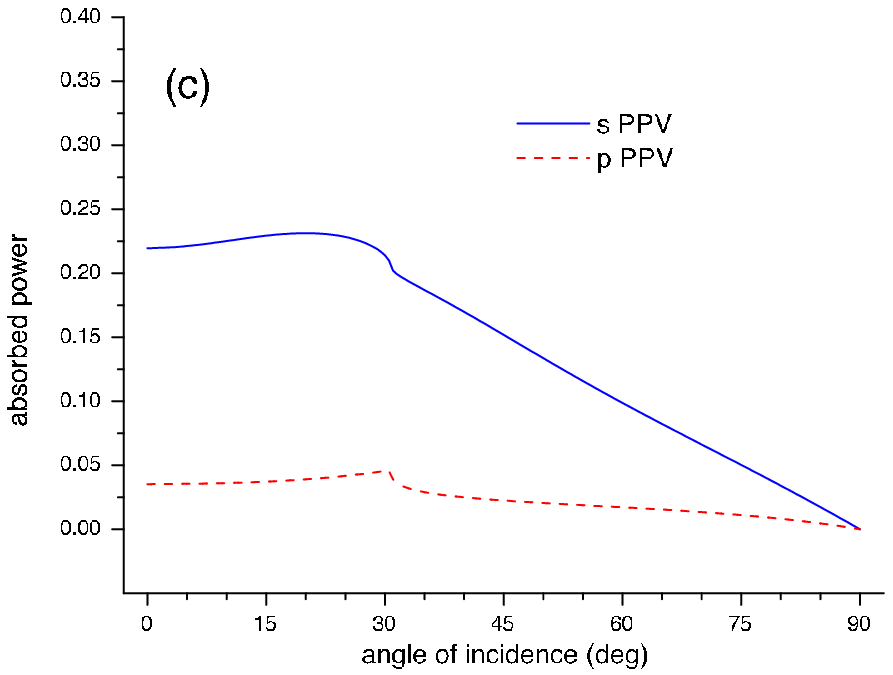}
\end{tabular}
\end{center} 
\caption[example]{ \label{surfwave4} Same as Figure \ref{surfwave3}, 
except that the refracting material is replaced by its PPV analog  ($\epsilon_2=1.8+i0.01,\,\mu_2=-1.5+i0.01$).}
 \end{figure}

For gratings with deep corrugations, the wavenumber of the surface wave can be appreciably 
different from the values in the absence of the corrugations, or the surface wave  can even be forbidden to propagate. 
This can be concluded from Figures \ref{surfwave3} and Figures \ref{surfwave4}, which were drawn
for the same parameters as for Figures \ref{surfwave1} and Figures \ref{surfwave2}, except that 
now $h/d=1$. Apparently, the $p$--polarized surface wave does still play a role in the diffraction 
by the NPV grating (Figure \ref{surfwave3}c), 
with a broad  absorption peak
at $\theta_0 \approx 10.5^\circ$, close to the value found for $h/d=0.07$
in Figure \ref{surfwave1}b.
But no resonant behavior can be seen in Figure \ref{surfwave4}, the refracting material then
being of the PPV type.
The Rayleigh--Wood anomaly determined by $\beta_{-1}^{(1)}=0$ is indicated in both figures at
$\theta_0 \approx 30.5^\circ$. 

\bigskip

\bigskip
 \bigskip
 
\begin{figure}[ht] 
\begin{center} 
\begin{tabular}{c}
\includegraphics[width=2.5in]{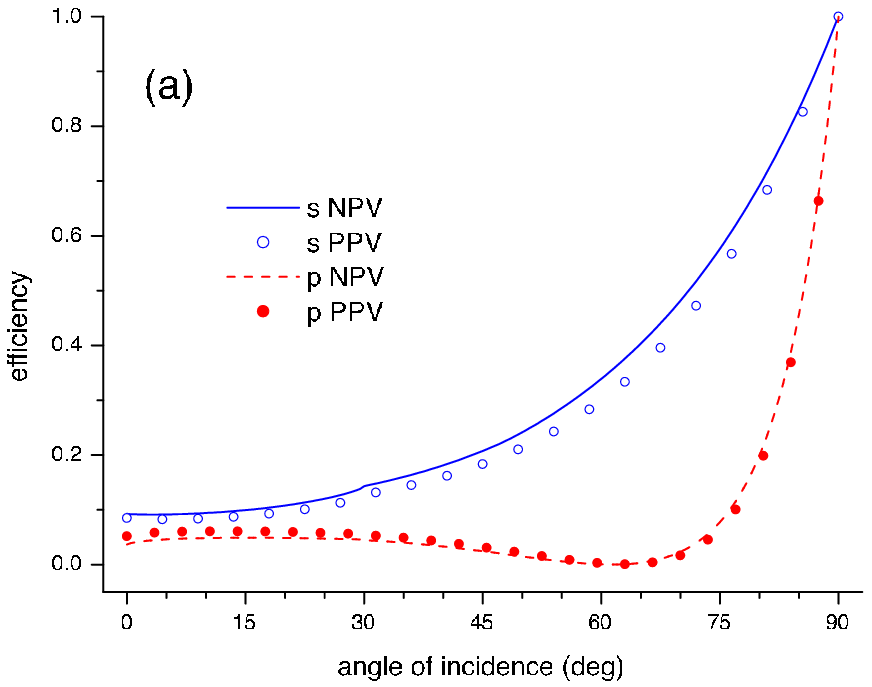} \hspace{0.3cm}
\includegraphics[width=2.5in]{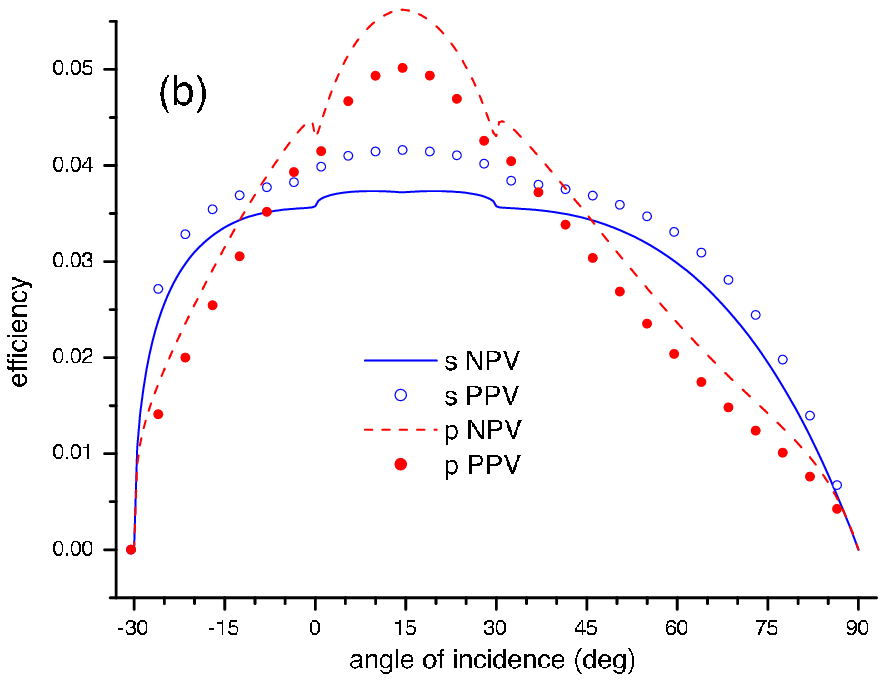} \\  \vspace{0.4cm}
\includegraphics[width=2.5in]{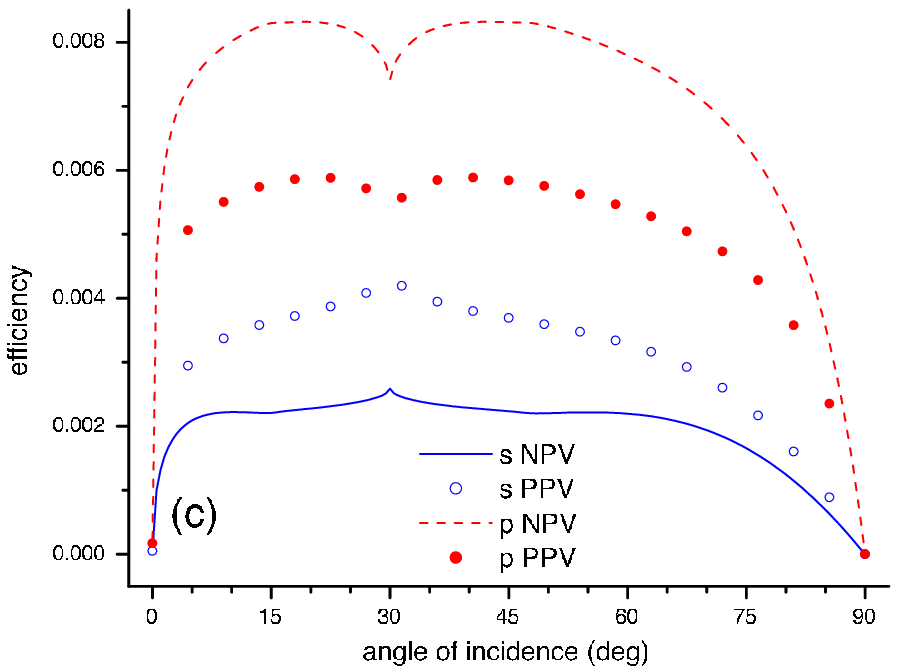} \hspace{0.3cm}
\includegraphics[width=2.5in]{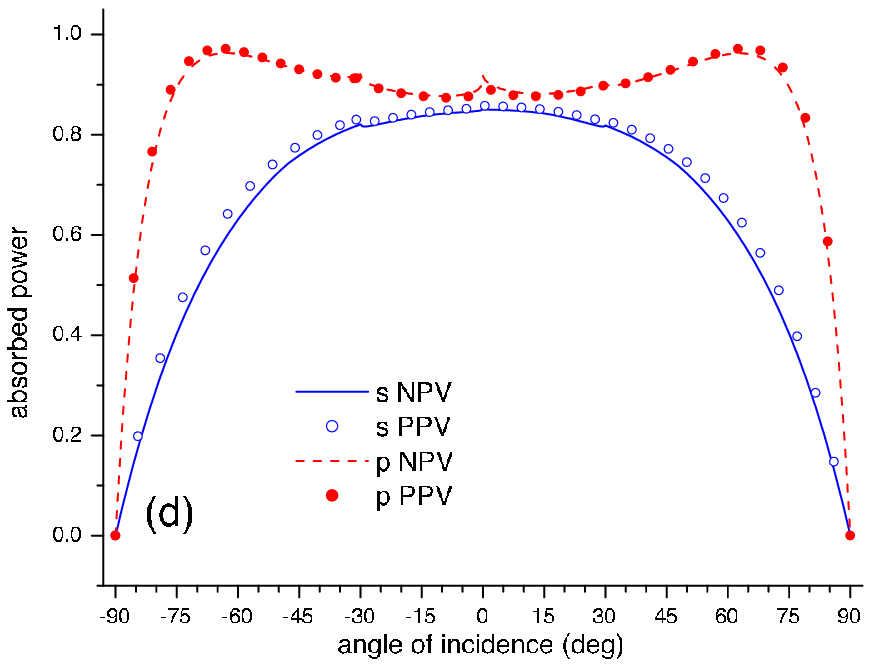} 
\end{tabular}
\end{center} 
\caption[example]{ \label{asym1} Diffraction efficiencies (a) $e_0^r$, (b) $ e_{-1}^r$, and 
(c) $e_{-2}^r$, and (d) normalized absorbed power $P_a$, as  functions of the incidence 
angle $\theta_0$, when $\omega d/c = 2\pi/0.5$. The corrugation shape is given by   (\ref{sum1}), 
with $h_1/d=0.12$, $h_2/d=0.078$, and $\gamma=\pi/2$. The refracting material is of either the PPV 
($\epsilon_2=6+i0.01,\,\mu_2=1+i0.01$) or the NPV ($\epsilon_2=-6+i0.01,\,\mu_2=-1+i0.01$)
type. Calculations were made for both the $s$-- and the $p$--polarization cases, 
using the C formalism with $N=29$.}
\end{figure}

\begin{center}{\em (d) Asymmetric corrugations}\end{center}
In order to illustrate effect of  the corrugation shape on the diffraction efficiencies,  we also considered asymmetric corrugations described by 
(\ref{sum1}).
Diffraction efficiencies were calculated for
 $h_1/d=0.12$, $h_2/d=0.078$ and $\gamma=\pi/2$, so that $\left({\rm max} g(x) -{\rm min} g(x)\right)/d \simeq 0.33 $.
 Calculated values of  $e_0^r$, $e_{-1}^r$, $e_{-2}^r$, and   $P_a$ as  functions of the incidence angle $\theta_0$ are plotted in Figure \ref{asym1}
for $\omega d/c = 2\pi/0.5$, when the 
refracting material is of either the PPV ($\epsilon_2=5+i0.01,\,\mu_2=1+i0.01$) or the NPV ($\epsilon_2=-5+i0.01,\,\mu_2=-1+i0.01$) type.
These plots were made for
 both the $s$-- and the $p$--polarization cases, using
 the C formalism with $N=29$. Clearly, application of the C formalism is not limited to 
 simple sinusoidal gratings. Additionally,
as in Sections IV (a) and (b), the differences between NPV and PPV gratings are easy to divine from 
 Figure \ref{asym1}, and Rayleigh--Wood anomalies are present therein.

\bigskip
\bigskip

\bigskip
\begin{figure}[ht] 
\begin{center} 
\begin{tabular}{c}
\includegraphics[width=2.5in]{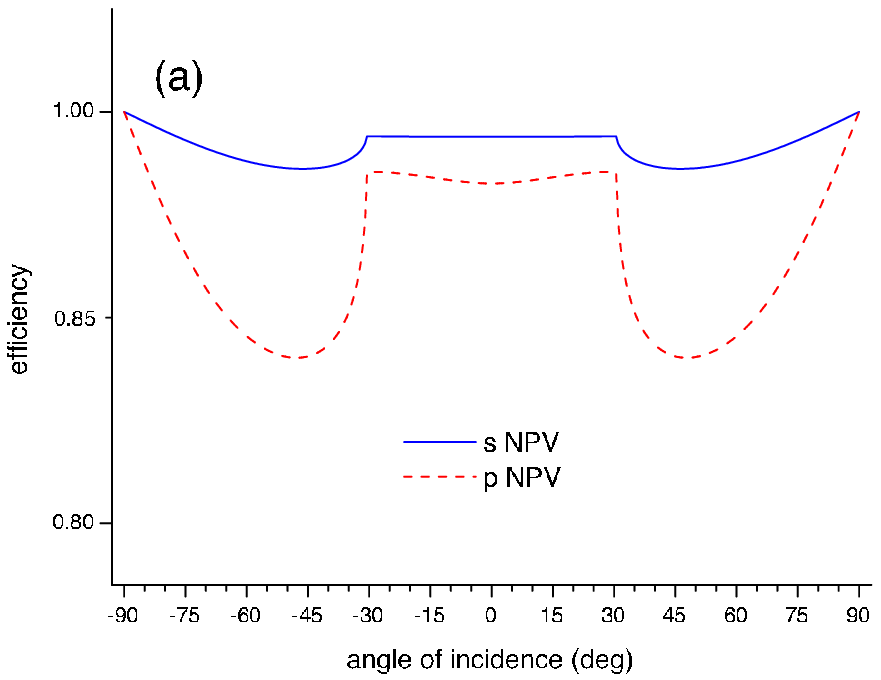} \hspace{0.3cm}
\includegraphics[width=2.5in]{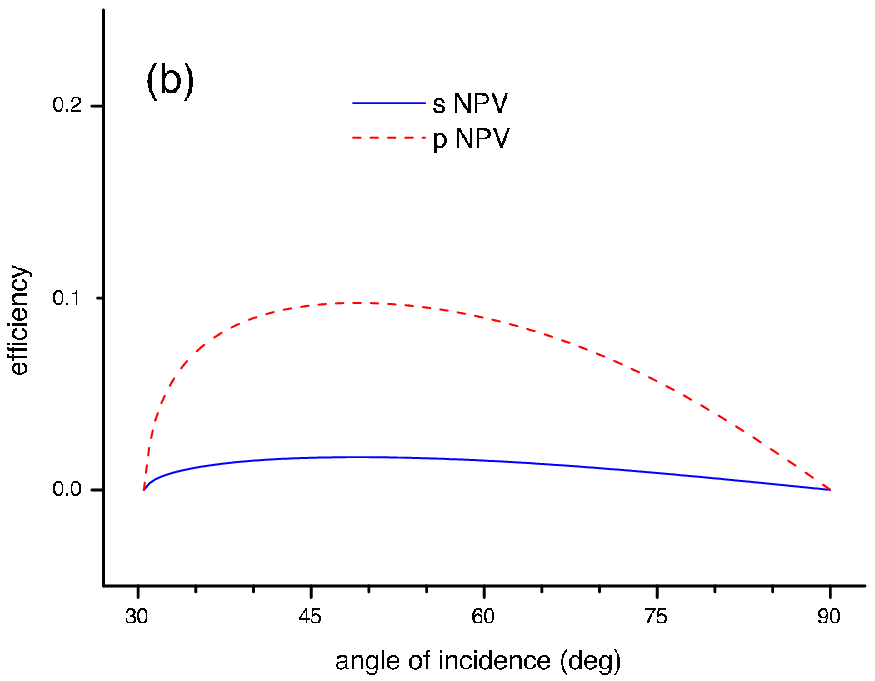} \\  \vspace{0.4cm}
\includegraphics[width=2.5in]{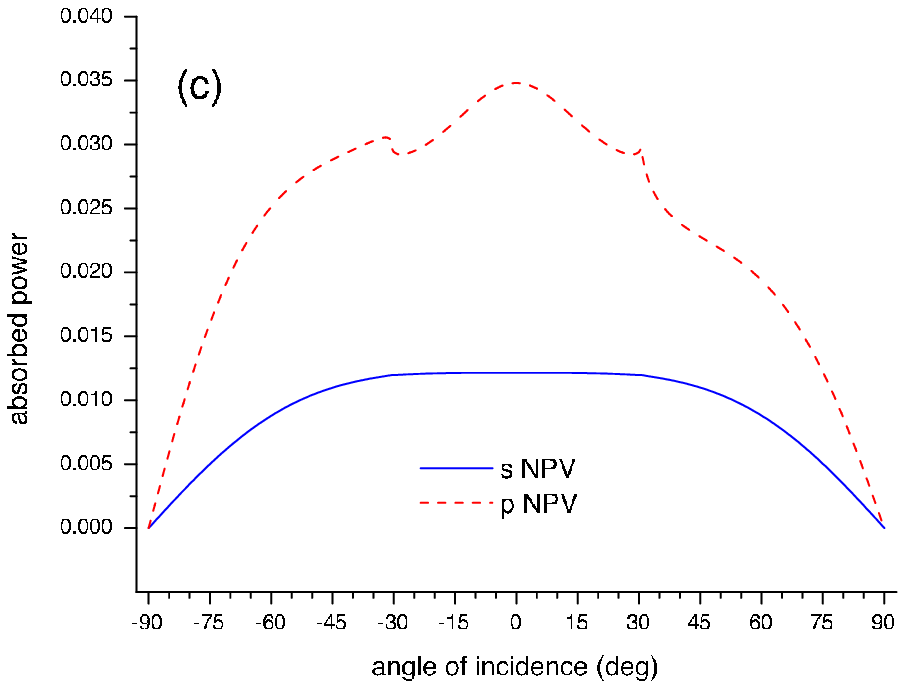}
\end{tabular}
\end{center} 
\caption[example]{ \label{asym2} Diffraction efficiencies (a) $e_0^r$ and (b )$e_{-1}^r$, and (c)
normalized absorbed power $P_a$  as 
functions of the incidence angle $\theta_0$, when $\omega d/c = 2\pi/1.51$. 
The corrugation shape is given by   (\ref{sum1}), 
with $h_1/d=0.04$, $h_2/d=0.026$, and $\gamma=\pi/2$. 
The refracting material is of the NPV ($\epsilon_2=-1.8+i0.01,\,\mu_2=1.5+i0.01$) type. 
Calculations were made for both the $s$-- and the $p$--polarization cases, using the C 
formalism with $N=29$.}
\end{figure}

\begin{figure}[ht] 
\begin{center} 
\begin{tabular}{c}
\includegraphics[width=2.5in]{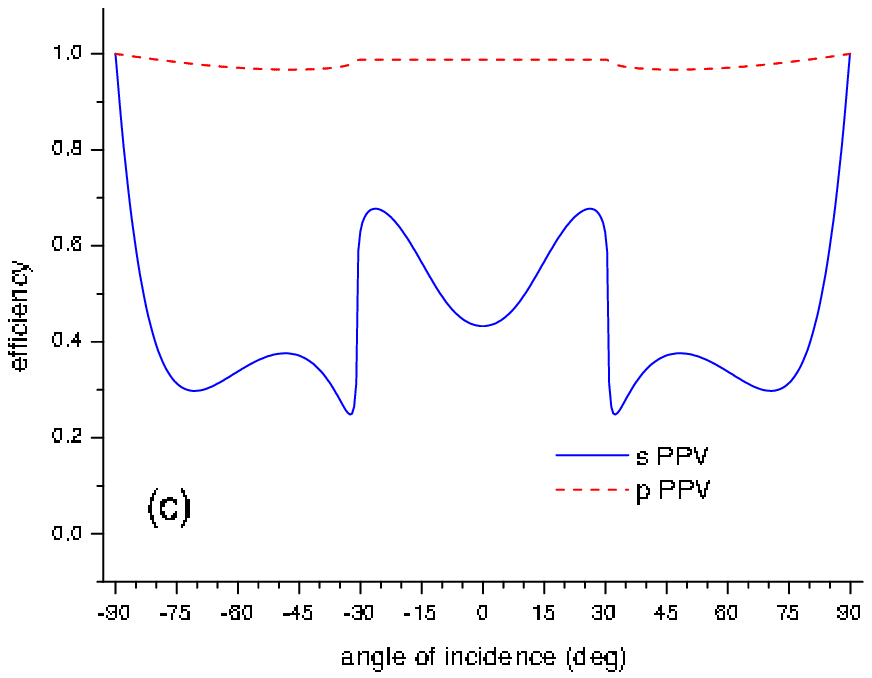} \hspace{0.3cm}
\includegraphics[width=2.5in]{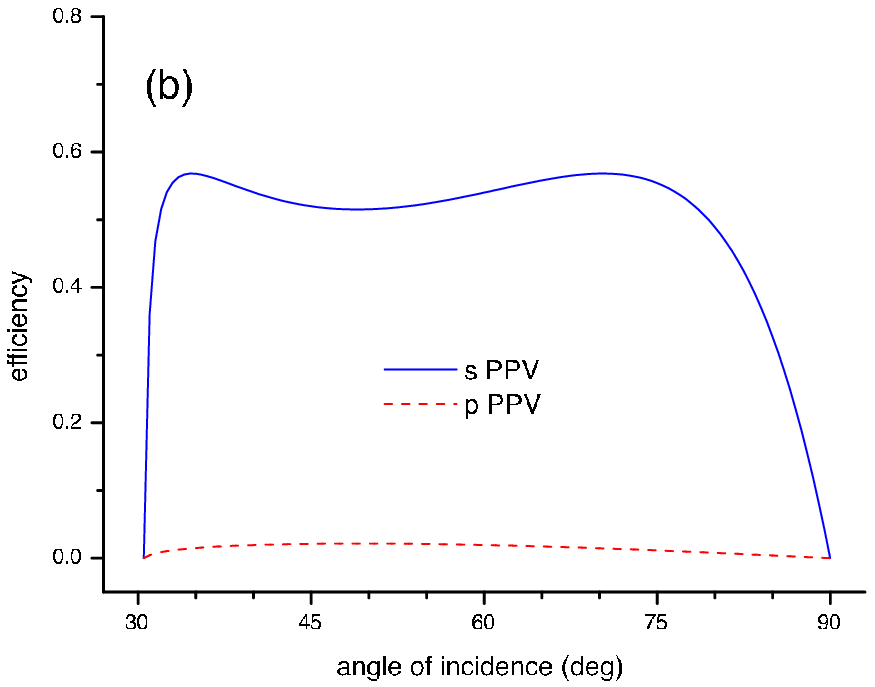} \\  \vspace{0.4cm}
\includegraphics[width=2.5in]{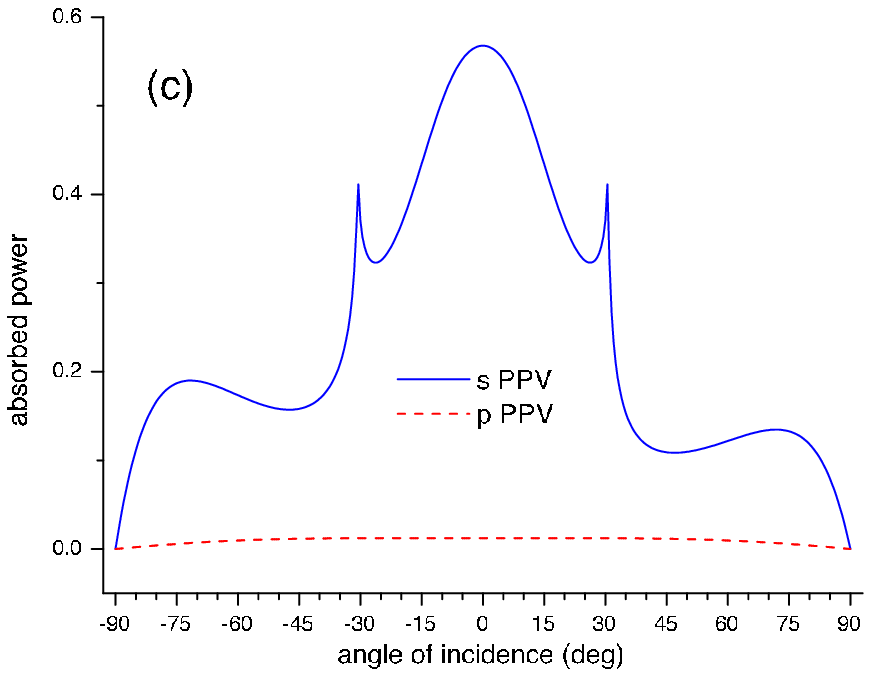}
\end{tabular}
\end{center} 
\caption[example]{ \label{asym3} Same as Figure \ref{asym2}, except that the refracting material is replaced by its PPV analog  ($\epsilon_2=1.8+i0.01,\,\mu_2=-1.5+i0.01$).}
 \end{figure}

The corrugation shape should affect surface wave propagation. This conjecture was verified when
the calcuations for Figures \ref{surfwave1} 
($\epsilon_2=-1.8+i0.01,\,\mu_2=1.5+i0.01$) and \ref{surfwave2}
($\epsilon_2=1.8+i0.01,\,\mu_2=-1.5+i0.01$) were repeated, but for the shape delineated
by (\ref{sum1}). Figures \ref{asym2} and \ref{asym3} were drawn for
 $h_1/d=0.04$, $h_2/d=0.026$ and $\gamma=\pi/2$. As $\left({\rm max} g(x) -{\rm min} g(x)\right)/d \simeq 0.11 $
 is rather small, the wavenumber of the surface wave should
 be predicted reasonably well by (\ref{surfs}) and (\ref{surfp}).
 But the
 introduction of a Fourier harmonic to 
a sinusoidal corrugation seems to change strongly the coupling between the surface and 
incident waves, both for NPV and for PPV materials, as can be observed from Figures 
\ref{asym2} and \ref{asym3}.

Finally, we must remark on a major difference and a major
similarity between Figures \ref{perturbresults}--\ref{surfwave4} on the
one hand and Figures \ref{asym1}--\ref{asym3} on the other. The corrugation shape is symmetric
for the former set of figures, but asymmetric for the latter. 
We see that  $P_a(\theta_0) = P_a(-\theta_0)$ for
symmetric corrugations, but $P_a(\theta_0) \ne P_a(-\theta_0)$ for asymmetric corrugations. However, whether or not the corrugations are symmetric, 
$e_0^r(\theta_0)=e_0^r(-\theta_0)$, 
as expected from reciprocity arguments (Petit 1980).

\bigskip
 \bigskip

\section{Concluding remarks}

In the foregoing sections, we extended Rayleigh's own method, a perturbative approach, and the C formalism to encompass
diffraction by surface--relief gratings made of made of isotropic, negative phase--velocity materials. This was enabled by 
carefully representing the field inside the refracting material. Numerical results for 
corrugations of both symmetric as well as asymmetric shapes were obtained, as also for both $s$-- and
$p$--polarized incident plane waves. We concluded that replacement of a PPV grating by its NPV analog affects only nonspecular
diffraction efficiencies when the corrugations are shallow. When the corrugations deepen,
the  specular diffraction efficiencies are also affected by the type of the refracting material. 

Surface wave propagation as well as the 
resonant excitation of surface waves also depends on whether the refracting material is of the NPV type or its PPV analog. 
Excitation of a surface wave through a grating plays
a fundamental role when  high selectivity is desired. 
In common PPV gratings, surface waves have been exploited for efficient conversion of $p$-- to $s$-- polarizations, or {\em vice versa\/}
 in conical mountings as well as to obtain enhanced nonlinear optical effects through the 
enhancement of surface fields usually associated with the resonant excitation of surface polaritons \cite{Boardman1982, Raether1988}. 
Surface waves play an important role in the concept of a 
perfect lens realized using a NPV material, since the field of an image, which can not be 
focused by a normal lens, can be transferred through a NPV layer by the excitation of surface waves at both 
of its boundaries \cite{Pendry2003, Shadrivovetal2004}. We expect that NPV slabs with periodically
corrugated boundaries combine both attributes.

\vspace{0.7cm}
\noindent {\bf Acknowledgments.}
R.A.D. acknowledges financial support from Consejo Nacional de 
Investigaciones Cient\'{\i}ficas
y T\'ecnicas (CONICET), Agencia Nacional de Promoci\'on 
Cient\'{\i}fica y Tecnol\'ogica (ANPCYT-BID 802/OC-AR03-04457) and
Universidad de Buenos Aires (UBA). A.L. acknowledges partial support 
from the Penn State Materials   Research Science and Engineering 
Center.

\end{document}